\documentclass[useAMS,usenatbib]{mnras}
\usepackage{graphics}
\usepackage{amssymb}
\usepackage{amsfonts}
\usepackage{wasysym}
\usepackage{epsfig}

\newcommand{\s}{\scriptsize}

\newcommand{\m}{\mathrm}
\title[The magnetosphere of TVLM\,513]
{
Probing the magnetosphere of the M8.5 dwarf TVLM\,513-46546 by modelling its auroral radio emission. Hint of star exoplanet interaction? 
}
\author[P. Leto et al.]
{P. Leto$^{1}$ \thanks{E-mail: pleto@oact.inaf.it},
C. Trigilio$^{1}$,
C. S. Buemi$^{1}$,
G. Umana$^{1}$,
A.~Ingallinera$^{1}$,
L. Cerrigone$^{2}$
\\
$^{1}$INAF - Osservatorio Astrofisico di Catania, Via S. Sofia 78, 95123 Catania, Italy\\
$^2$ASTRON, the Netherlands Institute for Radioastronomy, PO Box 2, 7990 AA Dwingeloo,The Netherlands
}
\begin{document}

\date{}

\pagerange{\pageref{firstpage}--\pageref{lastpage}} \pubyear{}

\maketitle

\label{firstpage}

\begin{abstract}
In this paper we simulate the cyclic circularly-polarised pulses of
the ultra-cool dwarf TVLM\,513-46546,
observed with the VLA at 4.88 and 8.44 GHz on May 2006,
{by using a 3D model of the auroral radio emission
from the stellar magnetosphere. 
During this epoch, the radio light curves are characterised by two pulses left-hand polarised at 4.88 GHz,
and one doubly-peaked (of opposite polarisations) pulse at 8.44 GHz.
To take into account the possible deviation from the dipolar symmetry of the stellar magnetic field topology,
the model described in this paper is also able 
to simulate the auroral radio
emission from a magnetosphere shaped like an offset-dipole.
To reproduce the timing and pattern of the observed pulses,
we explored the space of parameters controlling the auroral beaming pattern
and the geometry of the magnetosphere.
Through the analysis of the TVLM\,513-46546 auroral radio emission,
we derive some indications on the magnetospheric field topology 
that is able to simultaneously reproduce the timing and patterns of the auroral pulses measured at 4.88 and 8.44 GHz.
Each set of model solutions simulates two auroral pulses (singly or doubly peaked) per period.
To explain the presence of only one 8.44 GHz pulse per period, 
we analyse the case of auroral radio emission}
limited only 
to a magnetospheric sector activated by an external body, 
like the case of the interaction of Jupiter with its moons.
\end{abstract}


%________________________________________________________________
\section{Introduction}

The auroral radiation is a broad-band radio emission commonly observed 
in the magnetised planets of the solar system  \citep{zarka98}, originating in the auroral magnetospheric cavities
and interpreted in terms of
Electron Cyclotron Maser (ECM). 
The ECM is a coherent emission mechanism that amplifies 
the extraordinary magneto-ionic mode,
producing almost 100\% circularly polarised radiation at frequencies
close to the values that 
satisfy the cyclotron resonance condition $\nu=s \nu_{\mathrm B} / \gamma+k_{\parallel}v_{\parallel}$ \citep{melrose_dulk82}
(with $s$ the harmonic number; $\nu_{\mathrm B}=2.8 \times 10^{-3} B/{\m G}$ GHz the  local gyro-frequency;
$\gamma$ the Lorentz factor of the resonant electrons;  $k_{\parallel}$ and $v_{\parallel}$, the components 
of the wavevector $\mathbf{k}$ and of the electron velocity $\mathbf{v}$  parallel to the ambient magnetic field, respectively).
Such kind of amplification mechanism is
driven by an unstable electron energy distribution, like the loss-cone
\citep{wu_lee79,melrose_dulk82} or the horseshoe \citep{winglee_pritchett86} distributions.

The loss-cone distribution can be developed by a non-thermal
electron population propagating 
toward regions where the magnetic field strength increases. 
The electrons are accelerated far from the star due to magnetic reconnection.
Those with very low pitch angle 
(angle between the electron velocity and the local magnetic field vector)
can eventually impact on the stellar surface. As a result, the magnetically mirrored electron population will be deprived of these electrons,
giving rise to the loss-cone distribution.
In the weakly relativistic regime, the resonance condition for the loss-cone driven ECM mechanism 
can be written as: $\nu \geq s \nu_{\mathrm B}$, with the amplified radiation
beamed in a very thin hollow cone 
centred on the local magnetic field lines, and with a large half-aperture that is function of the frequency
and of the emitting electron speed  \citep*{hess_etal08}.

The electrons that move toward regions with increasing  magnetic field strength,
can also develop the unstable horseshoe distribution, if downward accelerated by a parallel electric field system.
Such unstable electron energy distribution is common for the Earth's Auroral Kilometric Radiation (AKR) \citep{ergun_etal00}.
The radiation amplified by the horseshoe-driven ECM mechanism
satisfies the cyclotron resonance condition for transverse emission ($k_{\parallel}v_{\parallel}=0$).
Then $\nu \leq s \nu_{\mathrm B}$
and the amplified radiation propagates perpendicularly to
the local magnetic field independently of the
emitted frequency or speed of the resonant emitting electrons \citep{hess_etal08}.

For all these cases the ECM amplification mechanism occurs within magnetospheric 
regions covering a wide range of values of magnetic field strength, which
 provides the broad-band observable auroral radio emission.
Moreover, in the case of auroral radio emission generated in thin magnetospheric cavities (laminar source model),
the overall radiation beam pattern will be strongly anisotropic. The emission beam is mainly directed along
the cavity wall perpendicularly to the local magnetic field vector \citep{louarn_lequeau96a,louarn_lequeau96b}.
Such strong anisotropy of the emission beam gives rise to a radio light-house effect of
the auroral radio emission.
The AKR angular beaming follows such laminar source model. After being emitted and passing through high-density regions, the radiation is refracted upwards \citep*{mutel_etal08,menietti_etal11}.

In the case of stars,
coherent pulses explained as auroral radio emission have been
observed in the hot chemically-peculiar and magnetic stars
CU\,Vir \citep{trigilio_etal00, trigilio_etal08, trigilio_etal11, ravi_etal10, lo_etal12}
and HD\,133880 \citep{chandra_etal15}.
{This coherent radio emission process has a counterpart also at X-ray wavelengths,
like the case of the hot magnetic star HR\,7355, which shows evidence
of X-ray aurorae \citep{leto_etal17}.}
These early type main sequence stars are characterised by a kGauss magnetic field with an overall dipolar topology. 
Their magnetic dipole axis is tilted with respect to the rotational axis
(oblique rotator model)  \citep{babcock49}.
Furthermore, at the bottom of the main sequence,
the signature of the auroral radio emission has been clearly identified in many very low mass stars
{and brown dwarfs, 
with spectral type ranging from M8 to T6.5
%UCDs showing  radio pulses: 
\citep*{berger02, %tvlm513 %2m 0036
burgasser_putman05, %denis1048
antonova_etal08, %2m 0746
hallinan_etal08, %lsr 1835
route_wolszczan12, % 2m 1047
route_wolszczan13, % 2m J1439+19 (tentative detection)
route_wolszczan16, %WISEP J112254.73+255021.5 
williams_etal15a, %2m J1314
burgasser_etal15, %2MASS J07200325-0846499  - WISE J072003.20Ð084651.2
kao_etal16}. % SIMP J01365662+0933473 - 2MASS 12373919+6526148 - 2M1043 - SDSS0423
%williams_etal17}. %WISEP J112254.73+255021.5
}

The stellar auroral radio emission detected in stars of both classes is characterised by strongly anisotropic beaming 
\citep*{trigilio_etal11,lo_etal12,lynch_etal15}. This is the case of the ECM emission that originates in a laminar source region.
The existence of a large-scale well-ordered magnetic field in 
the fast-rotator fully-convective M-type stars 
\citep{donati_etal06,donati_etal08,reiners_basri07,reiners_basri10,morin_etal08a,morin_etal08b,morin_etal10},
could explain the observed similarity 
of the auroral radio emission in these extreme classes of stars, like the case of hot magnetic stars.

In \citet{leto_etal16} (hereafter Paper I), we developed a 3D model able to simulate the
timing and the profile of the auroral pulses 
generated in  a dipolar-shaped laminar cavity. 
This model is a powerful tool to study the relation between the occurrence of auroral pulses  and the geometry of the stellar magnetosphere.
In this paper, we apply this model to reproduce the 
auroral radio emission properties of the well-studied M8.5 dwarf: TVLM\,513-46546 \citep{hallinan_etal07}.
To explain the features of its auroral radio emission, 
{we improved the auroral radio emission model to take into account the case of 
a magnetosphere shaped like an offset-dipole, and then
we, also,} considered the hypothesis that the unstable electron population 
could be partially originated by the interaction of an exoplanet with the stellar magnetosphere. 
This study was useful to give some constraints to the stellar geometry
and to the overall magnetic field topology.

{In Section~\ref{ucd_mcp}, %({\it THE RADIO EMISSION FROM UCDS AND MCPS}) 
we describe the main features at radio wavelengths commonly seen  in the class of the object in our study,
and compare these with hot magnetic stars.
Section~\ref{model} %({\it THE AURORAL RADIO EMISSION MODEL}) 
briefly describes the model of the stellar auroral radio emission
that was applied to the UCD TVLM\,513-46546 (Section~\ref{sim_tvlm}). %{\it Simulations of the TVLM 513-46546 auroral radio emission}).
The radio observations are presented in Section~\ref{sec_dati} %({\it VLA observations of TVLM 513-46546}), 
and the comparison with the simulations is described in Section~\ref{cso}. %({\it Comparison between simulations and observations}).
The case of auroral radio emission arising from a non-dipolarly shaped stellar magnetosphere is analysed in Section~\ref{non_dip} and
%({\it AURORAL RADIO EMISSION FROM A NON-DIPOLAR MAGNETOSPHERE}), 
details are given in the Appendix. %~\ref{offset_dip} and \ref{app}.
Constraints to the stellar magnetosphere % of TVLM\,513-46546, 
based on the analysis of the auroral radio emission features, 
are provided in Section~\ref{mag_tv}. % ({\it THE MAGNETOSPHERE OF TVLM 513-46546}).
%CITARE Section~\ref{unstab_electr}.
The considerations regarding the possible presence of an auroral radio emission component induced by a planet are discussed
in Section~\ref{planet}, % ({\it PLANET-INDUCED AURORAL EMISSION}),
while Section\,\ref{summ} summarises the results of this work. 
}

%________________________________________________
\section{The radio emission from UCDs and MCPs}
\label{ucd_mcp}

Very low-mass stars, late M type dwarfs, and the hot (L type) and cool (T type) brown dwarfs (commonly known as Ultra Cool Dwarfs, UCDs)
are very late-type main sequence stars at and below the hydrogen-burning limit.
In these kinds of objects, the classical magnetic-activity indicators (H${\alpha}$ and X-ray emission) fade away \citep{schmidt_etal07},
even though a number of these UCDs have been detected as non-thermal radio source
\citep*{berger_etal01,berger02,berger06,mclean_etal12,antonova_etal13,route_wolszczan13,kao_etal16,lynch_etal16,williams_etal17}.
If analysed in the context of the late-type stellar magnetic activity, the radio emission from  UCDs is peculiar. 
In fact, there is evidence that the radio emission from these objects violates
the empirical relation coupling the X-ray and radio luminosities of magnetically active stars 
($L_{\mathrm X} / L_{\mathrm {radio}} \approx 10^{15.5}$ Hz, \citealp{guedel_benz93,benz_guedel94}),
which is valid among stars distributed within a wide range of spectral classes (from F type to early M type).
In the UCD case, a decrease of the X-ray luminosity
is not related to a decrease of the radio luminosity. 
On the contrary, the radio luminosities of the UCDs are almost  constant, when detected 
\citep*{berger_etal10,williams_etal14,lynch_etal16}.
In spite of UCDs being characterised by absent or negligible X-ray coronae, the detection of their
radio emission  is indirect evidence that 
they can develop large magnetospheres filled with  non-thermal electrons and
extending up to tens of stellar radii \citep{nichols_etal12}.

The magnetic flux generation in
the fully convective stars at the bottom of the main sequence (spectral type earlier than M3)
is an open issue, since the classical $\alpha\Omega$ dynamo
characterising  solar-type stars and  early M dwarfs (M0--M3) does not work.
Nevertheless, the presence of magnetic fields at kGauss level
has been measured on the surface of a number of very late type UCDs \citep{reiners_basri07,reiners_basri10}.
These fields can be generated by other kinds of dynamo processes
working in fully convective stars, such as the  $\alpha ^2$-type dynamo \citep{kuker_rudiger99,chabrier_kuker06},
{or a process similar to the geodynamo operating on Jupiter and Earth \citep*{christensen_etal09}.}
The detection of fully polarised radio pulses at the microwave regime (a signature of stellar auroral radio emission)
is  further evidence that this kind of late-type dwarfs is characterised by magnetic fields at kGauss level \citep{kao_etal16}, 
being the emission frequency directly related to the local value of the magnetic field 
($\nu \approx \nu_{\mathrm B} \propto B$). 
The detection of the auroral radio emission 
requires the existence of fast electron beams precipitating toward the stellar surface,
%which has been proven to have a role as an ionising factor of the UCD atmospheres \citep{vorgul_helling16}.
which has a role as an ionising factor of the UCD atmospheres, as shown by \citet{vorgul_helling16}.

The study of the radio emission from  UCDs has a crucial role in understanding
the physical conditions of their magnetospheres,
but unfortunately the analysis of their radio emission is hampered by the weakness of the latter. 
The order of magnitude of the luminosity of a typical UCD 
detected at radio wavelengths is roughly in the range $10^{12}$--$10^{13}$ erg s$^{-1}$ Hz$^{-1}$ or less \citep{berger_etal10,mclean_etal12,williams_etal14,kao_etal16,gizis_etal16}. %,williams_etal17}.
 Taking into account also coherent events \citep{berger_etal08a}, this magnitude can  exceed
$10^{14}$ erg s$^{-1}$ Hz$^{-1}$, which anyway makes only a small sample observable. 

In many cases, observations were performed with poor spectral and temporal coverage.
Such lack of observational basis could prevent  us from theoretical-model approaches. 
In order to provide more input to understanding the radio behaviour of UCDs, we propose to take advantage of
the methods used in modelling the radio emission in hot Magnetic Chemically Peculiar (MCP) stars, 
which provide a unique possibility to study the plasma process in steady state within stable magnetic structures. 
MCPs are early-type main sequence stars (B/A spectral type), characterised by a mainly dipolar magnetic field at kGauss level.
A large fraction of the MCP stars are radio loud ($\approx 25\%$, \citealp{leone_etal94}) and their
radio emission is due to gyrosynchrotron from non-thermal electrons propagating 
in a large (tens of stellar radii) magnetospheric cavity.
In accordance with the oblique rotator model, the radio emission is also 
variable as a consequence of the stellar rotation \citep{leone91,leone_umana93}.

Despite the great difference between these two extreme spectral classes of stars,
many similarities are evident in the radio regime.
First of all, we notice the striking similarity of their radio light curves.
In particular, VLA measurements of the ultra cool dwarf 2MASS\,J13142039\,+\,1320011
highlight a sinusoidal flux-density modulation explained in the framework of the oblique rotator model \citep{mclean_etal11}.
This matches with a magnetic field topology resembling a tilted dipole.
The change of the magnetospheric projected area on the plane of the sky due to stellar rotation
explains the modulated radio emission.
In previous works, we proved that this scenario is able to reproduce the 
rotational modulation of the gyrosynchrotron emission from  MCP stars \citep{trigilio_etal04,leto_etal06}, 
which are oblique rotators. 
This similarity is reinforced by the detection   in both classes of stars of fully circularly polarised pulses, 
which are explained as stellar auroral radio emission.
Like in  MCPs,  the features of the microwave radiation from UCDs can be explained
in terms of auroral radio emission associated with the entire stellar magnetosphere \citep{williams_etal15a,william_berger15}. 
 
The early-type MCP stars are characterised by a magnetic field strength at kGauss level, like UCDs.
Although the behaviour of MCPs at radio wavelengths is similar to that of UCDs,
 their radio luminosity is about $10^4$ times higher (mean radio luminosity of the MCP stars 
$\approx 10^{17}$ erg s$^{-1}$ Hz$^{-1}$, \citealp*{drake_etal87,linsky_etal92}). 
The reason of such a huge difference  is easily related to the volumes of their magnetospheres.
The radii of the MCP stars are a few R$_{\odot}$, therefore
the magnetospheric volume of these stars can contain roughly $10^4$ magnetospheres of a typical UCD, 
which has a stellar radius of $\approx0.1$  R$_{\odot}$.

%============================================fig 1
\begin{figure}
\resizebox{\hsize}{!}{\includegraphics{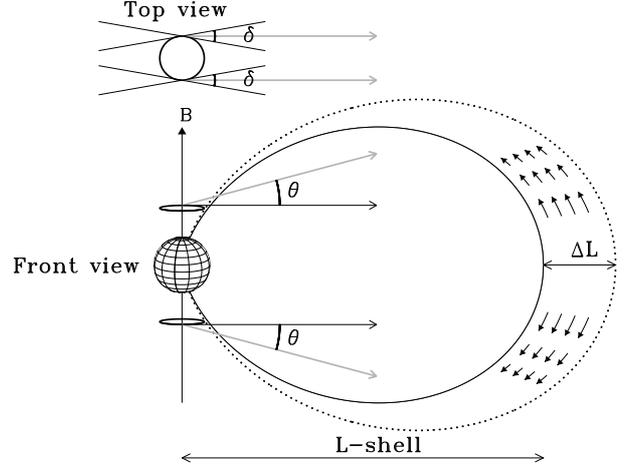}}
\caption{Cartoon showing the geometry of the stellar auroral radio emission.
The geometry of the emission beam pattern  (ray path deflection angle $\theta$ and opening angle $\delta$)
and the auroral ring are displayed. 
The grey arrows represent the direction of propagation of the rays.
The dipole magnetic field lines that identify the auroral cavity are pictured.
The small arrows directed toward the star represent the non-thermal electron population
that can develop the unstable energy distribution able to drive the ECM emission mechanism. }
\label{model_auroral}
\end{figure}
%============================================fig 1

The hot MCP stars have their magnetospheres dominated by a radiatively-driven stellar wind
extending for tens of stellar radii \citep{trigilio_etal04,leto_etal06},
whereas many UCDs are very fast rotators, with their magnetospheres dominated by rotation \citep{schrijver_09}. 
The role of the rotational speed to explain the UCD radio activity
has been highlighted by many authors \citep{berger_etal08b,mclean_etal12}.  
In  MCP-type stars, current sheets are formed by the magnetic field lines breaking near the Alfven radius\footnote{This radius defines the magnetospheric region where the wind kinetic pressure equals the magnetic one.}.
Similarly in fast-rotating UCDs, there are field-aligned current systems 
related to the co-rotation breakdown of the magnetospheric plasma,
which becomes centrifugally unstable at a distance of tens of stellar radii far from the surface \citep{nichols_etal12}.
In both cases, electrons can be accelerated up to relativistic energies.
Moreover, \citet{nichols_etal12} estimate
that the power of the electric current systems generated in the fast-rotating magnetospheres
is able to sustain the radio luminosity of the UCDs.
This non-thermal electron population can develop an unstable energy distribution 
able to drive the Electron Cyclotron Maser emission observed in some UCDs. These electrons eventually
propagate into the deeper magnetospheric regions radiating because of the incoherent
gyro-synchrotron mechanism.

The scenario described above is sketched in Fig.~\ref{model_auroral}.
The efficiency of the conversion mechanism of the rotationally-induced magnetospheric current systems 
in radio power decreases with the increasing
amount of plasma trapped in the magnetosphere, as a consequence of the reduction of the magnetospheric size
due to the larger plasma density \citep{nichols11,nichols_etal12}. 
The plasma density in the magnetosphere may not be stable over time, if it is provided by a mechanism external to the star, 
such as a volcanic planet.
This may also explain the episodic/seasonal behaviour of the auroral radio emission of the UCDs.
This is the case for Jupiter, where the auroral emission is directly affected by 
the density changes in Io's plasma torus due to the volcanic activity of this moon \citep{bonfond_etal12,yoneda_etal13}.
Therefore, by assuming an external volcanic body as the plasma source in the UCD radio-loud magnetospheres,
it is possible to explain the unstable auroral radio emission from this kind of objects.

\section{The auroral radio emission model}
\label{model}

{
%The auroral radio emission arising from the magnetised planets of the solar system
%is a well known phenomenon \citep{zarka98}.
The visibility of the planetary auroral radio emission and the corresponding spectral shape 
%frequency dependent features 
have been successfully modeled for Jupiter \citep{hess_etal08,hess_etal10,ray_hess08,cecconi_etal12},
% \citet{hess_etal08} simulation of the Jupiter dynamic specter observed from Earth \citep{quinnec_zarka98} 
%(Wind/Waves and the Nancay decameter array).
% \citet{cecconi_etal12} simulation of Jovian dynamic spectrum observed by the Planetary Radio Astronomy instrument onboard of the Voyager-2 spacecraft
and Saturn \citep{lamy_etal08}.
% paper \citep{ray_hess08}: Title=Modelling the Io-related DAM emission by modifying the beaming angle.
Following a similar approach, in Paper I}  %already used to simulate the planetary auroral radio emission, in Paper I}
we developed a 3D model to study the detectability conditions of the ECM pulses arising
from a laminar auroral cavity within a dipolar-shaped stellar magnetosphere. 
In particular, we proved how the study of the auroral radio-emission features
can give hints about  the geometry of the overall magnetospheric topology.
A section of the auroral radio emission model is sketched in Fig.~\ref{model_auroral}.

Due to the laminar structure of the auroral source region, the emission beam
pattern of the amplified radiation will be strongly anisotropic \citep{louarn_lequeau96a,louarn_lequeau96b}
(see the top view of the ring  in Fig.~\ref{model_auroral}).
It has been theoretically confirmed that the auroral radio emission is mainly amplified in the direction
tangent to the cavity wall \citep{speir_etal14}.
For a given stellar geometry (rotation axis inclination $i$ and tilt of the dipole axis $\beta$),
the model is able to reproduce the auroral radio emission pulse shape
and its phase occurrence as a function of the free parameters that control the emission beam pattern.
These parameters are the deflection angle $\theta$ and the beaming angle $\delta$.
The parameter $\theta$ is the angle between the wave vector $\mathbf{k}$ and the direction perpendicular to 
the magnetic field vector (roughly shown in the front view of the auroral source region  in Fig.~\ref{model_auroral}).
The angle $\delta$ quantifies the angular width of the emission beam pattern
tangentially directed along the auroral ring.

%___________________________________________________tab 1
\begin{table}
\caption[ ]{Stellar parameters of TVLM\,513-46546.}
\label{par_star}
\footnotesize
\begin{tabular}[]{lll}
\hline
\hline
\multicolumn{2}{l}{Fixed Parameters}\\
\hline
$R_{\ast}$ - stellar radius [R$_{\odot}$]                      & 0.103$^{\mathrm {a}}$              &                \\
$P_{\mathrm {rot}}$ - rotation period [hr]                       & 1.959574$^{\mathrm {b}}$              &                 \\
$i$ - rotation axis inclination [degree]                         & 70                               &               \\
$B_\mathrm{p}$ - polar magnetic field [Gauss]         & 3000$^{\mathrm {c}}$               &               \\
$\Delta \theta$ - hollow cone width  [degree]            &4$^{\mathrm {d}}$       &  \\
\hline
\multicolumn{1}{l}{Free Parameters} &Range & Step\\
\hline
$L$-shell [R$_{\ast}$]                                                      & 15--40$^{\mathrm {e}}$                  & 5                                  \\
$\theta$ - deflection angle [degree]                             & 0--25              & 2.5                  \\
$\delta$ - beaming angle [degree]                               & 5--30              & 5               \\
$\beta$ - magnetic axis obliquity [degree]                  & 0--355               & 5                  \\
\hline
\end{tabular}
\begin{list}{}{}
\item[References:]
(a) \citealp{hallinan_etal08};
(b) \citealp{wolszcan_route14};
(c) \citealp{hallinan_etal07};
(d) \citealp{berger_etal08a};
(e) \citealp{nichols_etal12}.
\end{list}
\end{table}
 %___________________________________________________tab 1

%============================================fig 2
\begin{figure*}
\resizebox{\hsize}{!}{\includegraphics{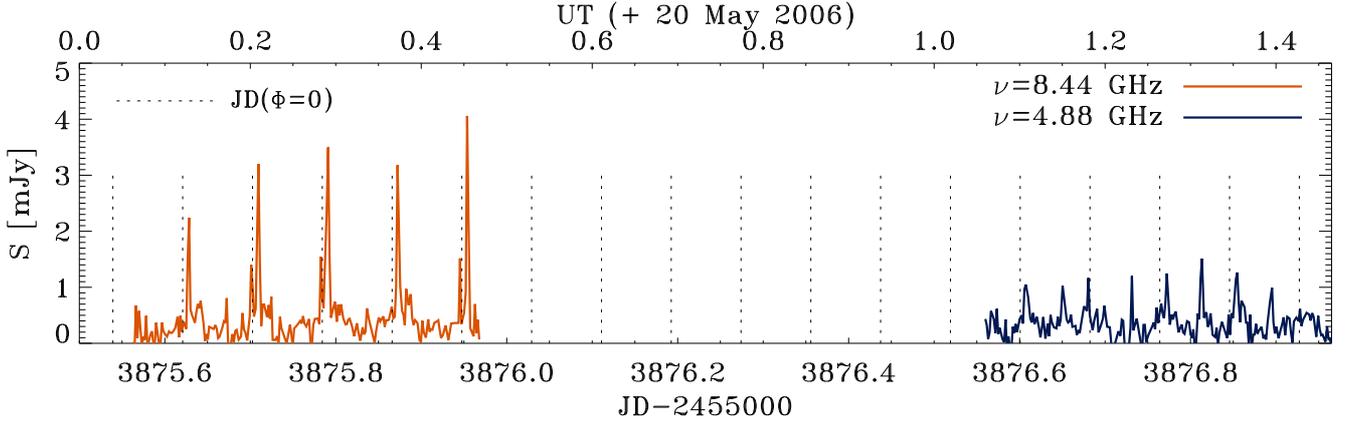}}
\caption{
Stokes I measurements of TVLM\,513-46546 performed with the VLA at 4.88 (C band) and 8.44 GHz (X band) 
during two consecutive days. 
Data already published by \citet{hallinan_etal07}.
Phase bin equal to 0.004. Average time 3 minutes.
}
\label{tvlm_obs}
\end{figure*}
%============================================fig 2

Given the frequency $\nu$ and the polar magnetic field strength $B_\mathrm{p}$,
the grid points where the condition $\nu=s \nu_{\mathrm B}$ is verified ($s$ harmonic number)
define auroral rings, northern and southern,
located above the polar caps, centred on the magnetic dipole axis, and parallel to the 
magnetic equatorial plane.
As a consequence of the magnetic field strength decreasing outwards (for a simple dipole, $B\propto r^{-3}$),
the frequency of the auroral radiation originating in rings close to the stellar surface is higher than 
the radiation emitted in farther rings.
The grid points of the magnetospheric cavity where the auroral radio emission
can take place are quantified  by the $L$-shell parameter
of the innermost magnetic field line 
(that is the distance to the star from the point where
the magnetic field line crosses the magnetic equator) and the shell thickness $\Delta L$;
these parameters are clearly pictured in Fig.~\ref{model_auroral}.

The model simulation of the auroral radio emission as a function of $\beta$, pointed out that 
a strict relation between the detectability of the ECM pulses 
and the ray path deflection ($\theta$) exists.
The analysis performed in Paper I showed the 
existence of two families of auroral light curves:
light curves having two doubly-peaked pulses per stellar rotation, with each single peak of opposite polarisation sign;
light curves characterised by two single peaks that can be left- or right-hand circularly polarised.
The doubly-peaked pulses point to the contributions of the two stellar hemispheres with opposite magnetic field orientation:
the auroral radio emission from the northern hemisphere is right-hand circularly polarised 
(Stokes V ($\mathrm{RCP}-\mathrm{LCP}$) $ > 0$), while
the southern emission is left-hand circularly polarised (Stokes V $< 0$).

Given the rotation axis inclination $i$, 
the phase separation between the peaks with opposite polarisation signs 
depends on the magnetic axis obliquity ($\beta$) and on the 
direction of propagation of the wave ($\theta$).
The auroral light curves showing only
singly-polarised pulses are related to those stellar geometries 
characterised by a lower tilt of the magnetic dipole axis.
In these cases, the effective magnetic field strength does not invert its sign as the star rotates
(magnetic curves without zeros)
and the auroral beam pattern arising from only one stellar hemisphere will cross the line of sight,
if characterised by the appropriate ray path deflection.
In general, the deflection angle $\theta$ controls 
the detectability and the phase location of the auroral pulses, while
the beam size of the auroral emission ($\delta$) is directly related to the phase width
of the single pulses. The main result obtained in Paper I
is that the phase location of the auroral radio pulses provides strong
constraints on the geometry of the stellar magnetospheres.

%___________________________________
 \subsection{Simulations of the TVLM\,513-46546 auroral radio emission}
 \label{sim_tvlm}

With the aim to investigate the overall physical conditions of the UCD magnetospheres,
we now apply the 3D model of the stellar auroral radio emission 
to a typical Ultra Cool Dwarf emitting coherent ECM pulses.
Among the members of this class of stars emitting auroral radio emission,
the M8.5-type dwarf TVLM\,513-46546 is the best studied in the radio regime,
from 327 MHz \citep{jaeger_etal11} to 95 GHz \citep{williams_etal15b}. This source is
characterised by quiescent (incoherent) and coherent radio emission.
Highly circularly polarised (both left- and right-hand) 
coherent pulses have been
observed  in C and X band with the {VLA, EVN, and Arecibo radio telescope in many epochs
\citep*{berger02,hallinan_etal06,hallinan_etal07,berger_etal08a,forbrich_berger09,doyle_etal10,yu_etal11,kuznetsov_etal12,
route_wolszczan13,wolszcan_route14,lynch_etal15,gawronski_etal16}},
but the star was also observed in quiescent status over 6 consecutive rotational periods \citep{osten_etal06}.
The rotation period of TVLM\,513-46546 was derived through
diagnostic methods based on infrared photometric monitoring
\citep{lane_etal07,littlefair_etal08,harding_etal13}. 
%The coherent pulses were phase-folded with the same period.
The coherent pulses were phase-folded with the same period.
%\textcolor{cyan}{\it The coherent pulses show the same period.}
This indicates that the pulsed ECM emission from TVLM\,513-46546
is a clear consequence of the stellar rotation \citep{hallinan_etal07}, like a light-house effect.

TVLM\,513-46546 has a high value of the projected rotational velocity
($v \sin i \approx 60 ~\mathrm{km\,s^{-1}}$; \citealp{mohanty_basri03}). Its
rotational period ($P_{\mathrm {rot}}$) is about 1.96 hr and is 
stable over a time scale close to a decade \citep{wolszcan_route14}.
As its radius, we take $R_{\ast} = 0.103$ R$_{\odot}$, the mean of the values given by \citealp{hallinan_etal08}. 
The combination of these parameters implies a very 
large inclination of the rotation axis,  $i =70^{\circ}$ (derived from the relation: $v \sin i = 2\pi R_{\ast} \sin i / P_{\mathrm {rot}}$).
To estimate the stellar magnetospheric size, we take advantage of
the analysis performed on this target by \citet{nichols_etal12}.
According to the authors, who adopt a polar field strength $B_\mathrm{p}=3000$ Gauss \citep{hallinan_etal07},
the magnetic latitude of the last closed field line lies in the
range 75$^{\circ}$--81$^{\circ}$. From the dipole field line equation, we know 
$r= L \cos ^2 \lambda $, where $r$ and  $ \lambda$ are the
distance to the centre of the star and the magnetic latitude, respectively.
The magnetic latitudes above correspond to a magnetospheric size in the range $\approx 15$--40 stellar radii.
The thickness of the auroral cavity ($\Delta L$) has been assumed about 40\% of the magnetospheric size, on average.
This value has been estimated roughly from  the magnetic-latitude range of the transition region where
the plasma coronation breakdown occurs ($\lesssim 2.5$ degree, \citealp{nichols_etal12}).
As highlighted in Paper I, this parameter has no effect on the simulated auroral radio emission features,
hence we fix its value to 40\% of the $L$-shell.

In the following simulation step, we fix $B_{\mathrm{p}}=3000$ Gauss.
The polar field strength directly affects the spatial location and  size of the auroral ring where
the ECM emission at a given frequency $\nu$ takes place.
As discussed in Paper I, the auroral radio emissions arising from a given magnetospheric cavity, 
but originating at different heights above the stellar surface, is characterised by almost indistinguishable ECM pulse profiles.
This prevents us from giving \textit{a priori} strong constraints on $B_{\mathrm p}$ and/or on the harmonic number $s$,
only on the basis of the auroral radio emission detection at a given frequency.
The assumed magnetic field strength corresponds to a polar gyro-frequency $\nu_{\mathrm B}=8.4$ GHz,
whereas at lower magnetic latitudes the auroral source location falls inside the star
(for a simple magnetic dipole the equatorial field strength is equal to half of the polar one).
To reproduce the auroral radiation at 8.4 GHz outside the stellar surface,
we assume that the ECM emission occurs at the second harmonic of the local gyro-frequency,
which is equivalent to assuming first-harmonic ECM emission from a dipolar magnetosphere with $B_{\mathrm p}=6000$ Gauss.
 
%============================================fig 3
\begin{figure}
\resizebox{\hsize}{!}{\includegraphics{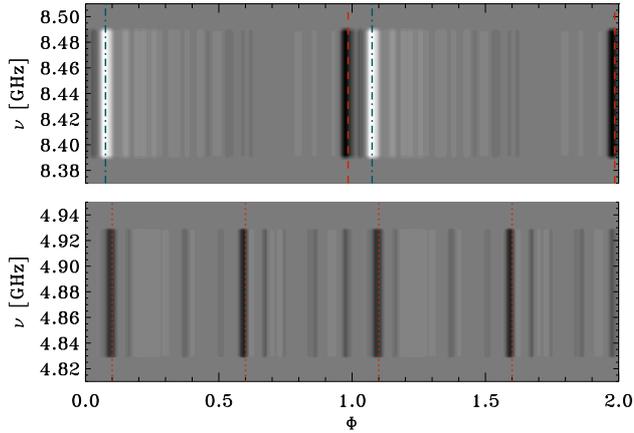}}
\caption{Phase folded light curve of TVLM\,513-46546 for the circularly polarised flux (Stokes V).
Bottom panel: C band,  top panel: X band. Phase bin equal to 0.004.
}
\label{tvlm_dic_spec}
\end{figure}
%============================================fig 3

In the model simulation performed here, we take
the angles $\theta$ and $\delta$ that control the auroral radio emission beaming as free parameters.
The parameter  $\Delta \theta$ is the thickness of the conical sheet where
the radiation generated by the elementary ECM-amplification process
 is beamed. This is related to the velocity of the unstable electron population as follows:
$\Delta \theta \approx v/c$ \citep{melrose_dulk82}. The energy of the emitting electron (about 1 keV) was
estimated on the basis of the TVLM\,513-46546 X-ray measurements \citep{berger_etal08a}.
The above electron energy corresponds to an emitting conical sheet $\approx 4^{\circ}$ thick.
The last  free parameter is the tilt angle $\beta$ of the magnetic dipole axis with respect to the rotation axis.

The model free parameters,
the adopted ranges of values, and the corresponding simulation step are summarised in Table~\ref{par_star},
for a total of $\approx 2.5\times 10^4$ simulated auroral radio light curves.
To properly sample the thin auroral cavity close to the stellar surface,
the sampling step was taken equal to 0.0225~R$_{\ast}$.

%============================================fig 4
\begin{figure*}
\resizebox{\hsize}{!}{\includegraphics{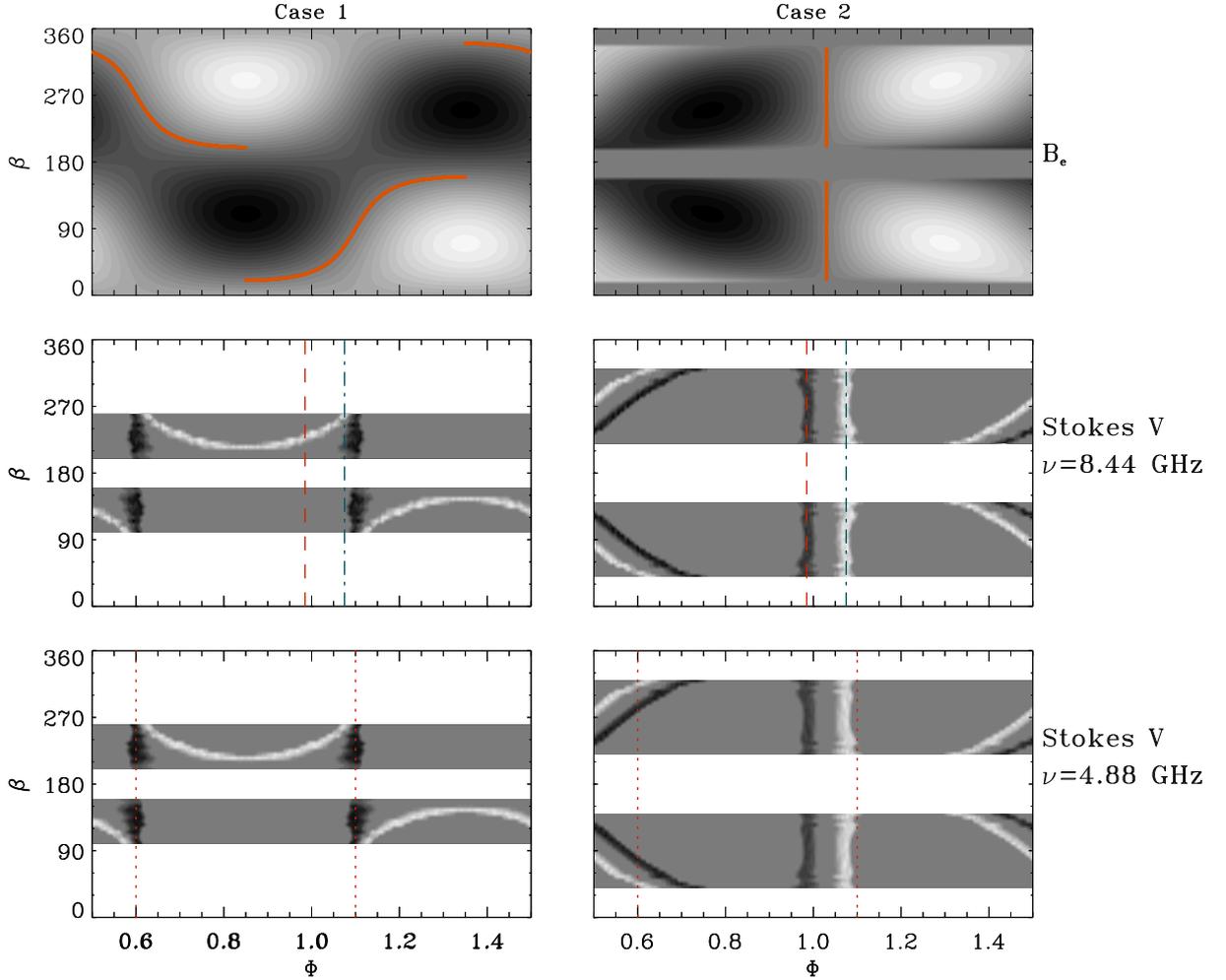}}
\caption{
Dynamic synthetic auroral light curves for the C ({$\nu=4.88$ GHz,} bottom panels) and X band  
({$\nu=8.44$ GHz,} middle panels) obtained as a function
of the angle $\beta$.
The dotted vertical lines locate the phase occurrence of the 4.88 GHz auroral pulses.
The left- and right-handed circularly polarised
pulse components at 8.44 GHz are represented with  dashed and  dot-dashed lines, respectively.
The variation of the effective magnetic curve is displayed in the top panels.
For the explanation of the 2 cases analysed see the text.
}
\label{tvlm_simulaz}
\end{figure*}
%============================================fig 4

%___________________________________________________tab 2
\begin{table*}
\caption[ ]{Model solutions.}
\label{sol_simul}
\footnotesize
\begin{tabular}[]{cc c ccc c cccr}
\hline
\hline
 & & &\multicolumn{3}{c}{Case 1: constraints from {$\nu=4.88$ GHz}} & &\multicolumn{4}{c}{Case 2: constraints from {$\nu=8.44$ GHz}}\\
%\cline{3-5}
% \cline{7-10} 
\cline{4-6}
 \cline{8-11} 
 $\beta$ &{$|\beta|$} & &$\theta$  &$\delta$ & Pulses &  &$\theta$ &$\delta$ &Pulses &$\Delta \Phi$ \\

[degree]  & {[degree]}    &     &[degree]        &[degree] & &            &[degree]        &[degree] & &\\
%  \cline{1-1}
%\cline{3-5}
% \cline{7-10} 

  \cline{1-2}
\cline{4-6}
 \cline{8-11} 
 
% \cline{12-12}

\s 40  &{\s 40}     &       &-                      &-        &-              &   & \s 10.0        &\s 15      &\s RCP+LCP   &\s  $-0.001$ \\
\s 45   &{\s 45}    &        &-                      &-        &-              &   & \s 10.0        &\s 15      &\s RCP+LCP    &\s   0.010 \\
\s 50   &{\s 50}    &        &-                      &-        &-              &   & \s 12.5        &\s 15      &\s RCP+LCP    &\s   0.020 \\
\s 55   &{\s 55}    &        &-                      &-        &-              &   & \s 12.5        &\s 20      &\s RCP+LCP    &\s   0.029 \\
\s 60   &{\s 60}    &        &-                      &-        &-              &   & \s 15.0        &\s 20      &\s RCP+LCP    &\s   0.036 \\
\s 65   &{\s 65}    &        &-                      &-        &-              &   & \s 15.0        &\s 20      &\s RCP+LCP    &\s   0.043 \\
\s 70   &{\s 70}    &        &-                      &-        &-              &   & \s 15.0        &\s 20      &\s RCP+LCP    &\s   0.049 \\
\s 75   &{\s 75}    &        &-                      &-        &-              &   & \s 15.0        &\s 20      &\s RCP+LCP    &\s   0.054 \\
\s 80   &{\s 80}    &        &-                      &-        &-              &   & \s 15.0        &\s 20      &\s RCP+LCP    &\s   0.060 \\
\s 85   &{\s 85}    &        &-                      &-        &-              &   & \s 15.0        &\s 20      &\s RCP+LCP    &\s   0.065 \\
\s 90   &{\s 90}    &        &-                      &-        &-              &   & \s 15.0        &\s 20      &\s RCP+LCP    &\s   0.070 \\
\s 95   &{\s 85}    &       &-                      &-        &-              &   & \s 15.0        &\s 20      &\s RCP+LCP    &\s   0.075 \\
\s 100 &{\s 80}     &       & \s ~5.0               &\s 10     &\s   RCP+LCP   &   & \s 15.0        &\s 20      &\s RCP+LCP    &\s   0.080 \\
\s 105 &{\s 75}     &       & \s ~5.0               &\s 10     &\s   RCP+LCP   &   & \s 15.0        &\s 20      &\s RCP+LCP    &\s   0.086 \\
\s 110 &{\s 70}     &       & \s ~7.5               &\s 10     &\s   RCP+LCP   &   & \s 15.0        &\s 20      &\s RCP+LCP    &\s   0.091 \\
\s 115 &{\s 65}     &       & \s ~7.5               &\s 15    &\s   RCP+LCP   &   & \s 15.0        &\s 20      &\s RCP+LCP    &\s   0.097 \\
\s $120^{\dag}$  &{\s 60}     &       & \s 10.0               &\s 15    &\s   RCP+LCP   &   & \s 15.0        &\s 20      &\s RCP+LCP    &\s   0.104 \\
\s 125  &{\s 55}   &       & \s 12.5               &\s 20    &\s   RCP+LCP   &   & \s 12.5        &\s 20      &\s RCP+LCP    &\s   0.111 \\
\s 130  &{\s 50}    &       & \s 12.5               &\s 20    &\s   RCP+LCP   &   & \s 12.5        &\s 15      &\s RCP+LCP    &\s   0.119 \\
\s 135  &{\s 45}    &       & \s 15.0               &\s 20    &\s   RCP+LCP   &   & \s 10.0        &\s 15      &\s RCP+LCP    &\s   0.129 \\
\s 140  &{\s 40}    &       & \s 15.0               &\s 20    &\s   RCP+LCP   &   & \s 10.0        &\s 15      &\s RCP+LCP    &\s   0.141 \\
\s 145  &{\s 35}    &       & \s 15.0               &\s 20    &\s   RCP+LCP   &   & -        &\s -        &\s -     &\s  -   \\
\s 150  &{\s 30}    &       & \s 17.5               &\s 20    &\s    LCP      &   & -         &\s -        &\s -     &\s   -   \\
\s 155  &{\s 25}    &       & \s 20.0               &\s 20    &\s    LCP      &   &-               &-          &-             &-              \\
\s 160  &{\s 20}    &       & \s 20.0               &\s 20    &\s    LCP      &   &-               &-          &-             &-              \\

\s 200  &{\s 20}    &       & \s 20.0               &\s 20    &\s    LCP      &   &-               &-          &-            &-              \\
\s 205  &{\s 25}    &       & \s 20.0               &\s 20    &\s    LCP      &   &-               &-          &-            &-              \\
\s 210  &{\s 30}    &       & \s 17.5               &\s 20    &\s    LCP      &   & -              &-          &-            &-                \\
\s 215  &{\s 35}    &       & \s 15.0               &\s 20    &\s  RCP+LCP    &   & -         &-        &-             &-  \\
\s 220  &{\s 40}    &       & \s 15.0               &\s 20    &\s  RCP+LCP    &   & \s 10.0        &\s 15      &\s RCP+LCP    &\s   $-0.359$ \\
\s 225  &{\s 45}    &       & \s 15.0               &\s 20    &\s  RCP+LCP    &   & \s 10.0        &\s 15      &\s RCP+LCP     &\s   $-0.371$ \\
\s 230  &{\s 50}    &       & \s 12.5               &\s 20    &\s  RCP+LCP    &   & \s 12.5        &\s 15      &\s RCP+LCP    &\s   $-0.381$ \\
\s 235  &{\s 55}    &       & \s 12.5               &\s 20    &\s  RCP+LCP    &   & \s 12.5        &\s 20      &\s RCP+LCP    &\s   $-0.389$ \\
\s 240  &{\s 60}    &       & \s 10.0               &\s 15    &\s  RCP+LCP    &   & \s 15.0        &\s 20      &\s RCP+LCP    &\s   $-0.396$ \\
\s 245  &{\s 65}    &       & \s ~7.5               &\s 15    &\s  RCP+LCP    &   & \s 15.0        &\s 20      &\s RCP+LCP    &\s   $-0.403$ \\
\s 250  &{\s 70}    &       & \s ~7.5               &\s 10     &\s  RCP+LCP    &   & \s 15.0        &\s 20      &\s RCP+LCP    &\s   $-0.409$ \\
\s 255  &{\s 75}    &       & \s ~5.0               &\s 10     &\s  RCP+LCP    &   & \s 15.0        &\s 20      &\s RCP+LCP    &\s   $-0.414$ \\
\s 260  &{\s 80}    &       & \s ~5.0               &\s 10     &\s  RCP+LCP    &   & \s 15.0        &\s 20      &\s RCP+LCP    &\s   $-0.420$ \\
\s 265  &{\s 85}    &       &-                      &-        &-              &   & \s 15.0        &\s 20      &\s RCP+LCP    &\s   $-0.425$ \\
\s 270  &{\s 90}    &       &-                      &-        &-              &   & \s 15.0        &\s 20      &\s RCP+LCP    &\s   $-0.430$ \\
\s 275  &{\s 85}    &       &-                      &-        &-              &   & \s 15.0        &\s 20      &\s RCP+LCP    &\s   $-0.435$ \\
\s 280  &{\s 80}    &       &-                      &-        &-              &   & \s 15.0        &\s 20      &\s RCP+LCP    &\s   $-0.440$ \\
\s 285  &{\s 75}    &       &-                      &-        &-              &   & \s 15.0        &\s 20      &\s RCP+LCP    &\s   $-0.446$ \\
\s 290  &{\s 70}    &       &-                      &-        &-              &   & \s 15.0        &\s 20      &\s RCP+LCP    &\s   $-0.451$ \\
\s 295  &{\s 65}    &       &-                      &-        &-              &   & \s 15.0        &\s 20      &\s RCP+LCP    &\s   $-0.457$ \\
\s 300  &{\s 60}    &       &-                      &-        &-              &   & \s 15.0        &\s 20      &\s RCP+LCP    &\s   $-0.464$ \\
\s 305  &{\s 55}    &       &-                      &-        &-              &   & \s 12.5        &\s 20      &\s RCP+LCP    &\s   $-0.471$ \\
\s 310  &{\s 50}    &       &-                      &-        &-              &   & \s 12.5        &\s 15      &\s RCP+LCP    &\s   $-0.480$ \\
\s 315  &{\s 45}    &       &-                      &-        &-              &   & \s 10.0        &\s 15      &\s RCP+LCP    &\s   $-0.490$ \\
\s 320  &{\s 40}    &       &-                      &-        &-              &   & \s 10.0        &\s 15      &\s RCP+LCP    &\s   $-0.499$ \\
\hline
\end{tabular}
\begin{list}{}{}
%\item[] {{$^{\dag}$ Parameters adopted for the simulations of the X band performed in Sec.~\ref{planet}, {\bf and for the simulation of the dynamic spectrum performed in Sec.~\ref{size_depend}} }}
\item[] {{$^{\dag}$ Parameters adopted for the simulations of the dynamic spectra performed in Sec.~\ref{non_dip},
%and in the Appendix~\ref{size_depend},
and for the simulations of the X band ($\nu=8.44$ GHz) light curves performed in Sec.~\ref{planet}}}
%of the X band performed in Sec.~\ref{planet}, and for the simulation of the dynamic spectrum performed in Sec.~\ref{size_depend} }}
\end{list}
\end{table*}
%____________________________________________________________

%__________________________________
\subsection{VLA observations of TVLM\,513-46546}
\label{sec_dati}

%============================================fig 6
%\begin{figure*}
%\resizebox{\hsize}{!}{\includegraphics{case1_dx.eps}}
%\resizebox{\hsize}{!}{\includegraphics{case1_dy.eps}}
%\resizebox{\hsize}{!}{\includegraphics{case1_dz.eps}}
%\caption{\bf cippi cippi bau bau cippi cippi bau bau cippi cippi bau bau cippi cippi bau bau cippi cippi bau bau cippi cippi bau bau cippi cippi bau bau cippi cippi bau bau cippi cippi bau bau cippi cippi bau bau cippi cippi bau bau cippi cippi bau bau.  
%}
%\label{case1_4pole} 
%\end{figure*}
%============================================fig 6

To identify the set of free parameters able to reproduce the TVLM\,513-46546 auroral radio emission,
the model simulations
must be compared with real multi-wavelength observations of this target.
As previously highlighted, the radio emission features of the UCDs 
are not stable within all the observing epochs.
With the aim to minimise the source variability,
we selected a data set where the source TVLM\,513-46546
was characterised by an almost stable auroral radio emission 
lasting for several stellar rotation periods.
The observing epoch here analysed refers to the VLA\footnote{The Very Large Array is a facility of the National Radio
Astronomy Observatory which is operated by Associated Universities, Inc.
under cooperative agreement with the National Science Foundation.} measurements 
of TVLM\,513-46546 performed on May 2006 at 8.44 ($\approx 9.75$ hours on May 20th) 
and 4.88 GHz ($\approx 9.75$ hours on May 21st), within a bandwidth 100 MHz wide, 
already published by \citet{hallinan_etal07}. 
These data have been re-analysed using the standard procedures of the Astronomical Image Processing System (AIPS).
The observed light curves for the Stokes I in C (4.88 GHz) and X (8.44 GHz) bands are shown in Fig.~\ref{tvlm_obs}.

We analyse the present observing epoch because TVLM\,513-46546 
has passed through a stable active phase,
showing ECM pulses at every rotation period.
This could help us to simulate the true modulation effect of its auroral radio emission 
without suffering from the intrinsic source variability.
The measurements performed on May 20th at 8.44 GHz lasted about 5 consecutive rotation periods and
periodic doubly-peaked pulses are clearly detected. 
Periodic pulses have been clearly detected also at 4.88 GHz, 
during the other 5 consecutive rotation periods observed on May 21st, but with a lower peak level (see Fig.~\ref{tvlm_obs}).

These VLA measurements were phase folded using the ephemeris given by \citet{wolszcan_route14}:
\\
\\
\noindent
$\mathrm{MJD}=2\,456\,188.65+0.0816489 E  ~~\mathrm{[days]}$.
\\
\\
\noindent
The phase-folded light curves for the circularly polarised flux density (Stokes V), 
are shown in Fig.~\ref{tvlm_dic_spec} for both C and X bands.
As discussed by \citet{hallinan_etal07}, the doubly-peaked pulses detected in X band
have opposite polarisation signs (see Fig.~\ref{tvlm_dic_spec} top panel),
and lie in the range of phases $\approx 0.95$--1.1.
The light curve for the C band 
is instead characterised by two pulses per spin period that are left-hand circularly polarised  
($\Phi \approx 0.1$ and $\Phi \approx 0.6$, Fig.~\ref{tvlm_dic_spec} bottom panel).
Both pulses have a similar phase duration of about 5\% of the rotational period.

%______________________________
\subsection{Comparison between simulations and observations}
\label{cso}

In the next step, we compare the model simulations with the observations, assuming that 
the source is stable within the analysed epoch. 
The differences between the two observing bands are
assumed as an intrinsic frequency-dependent effect.
%{\it Even though the two bands are not observed simultaneously,
%this is a fully sensible hypothesis, also supported 
%by the almost simultaneous C and X broad-band observations performed with the
%Karl G. Jansky Very Large Array (JVLA)
%during other active epochs of TVLM\,513-46546, 
%which display very similar behaviour \citep{lynch_etal15}.}

The analysis of the auroral radio emission from TVLM\,513-46546 
aims at achieving  a deep knowledge of its magnetosphere.
First of all, we noticed that the size of the auroral cavity has null effect
on the simulated pulses, confirming the results obtained in Paper I. 
Above $\approx 10$ R$_{\ast}$ the $L$-shell parameter does not affect the phase location and the shape of  the ECM pulses.
The features of the auroral radio emission from TVLM\,513-46546 (described in Sec.~\ref{sec_dati}) indicate
that the magnetic curve extremes should be located at intermediate phases between the two pulses observed in C band.
On the other hand, the doubly-peaked structure (with single peaks of opposite polarisation)
detected in the X band gives a clear indication that it is sampling the rotational transition 
from the southern to the northern stellar hemisphere. In fact, each stellar hemisphere
contributes to the auroral radio emission with radiation of opposite polarisation.
The null  of the magnetic curve will then be falling at an intermediate phase between
the pulses of opposite signs detected in X band.

%============================================fig 6
\begin{figure*}
\resizebox{\hsize}{!}{\includegraphics{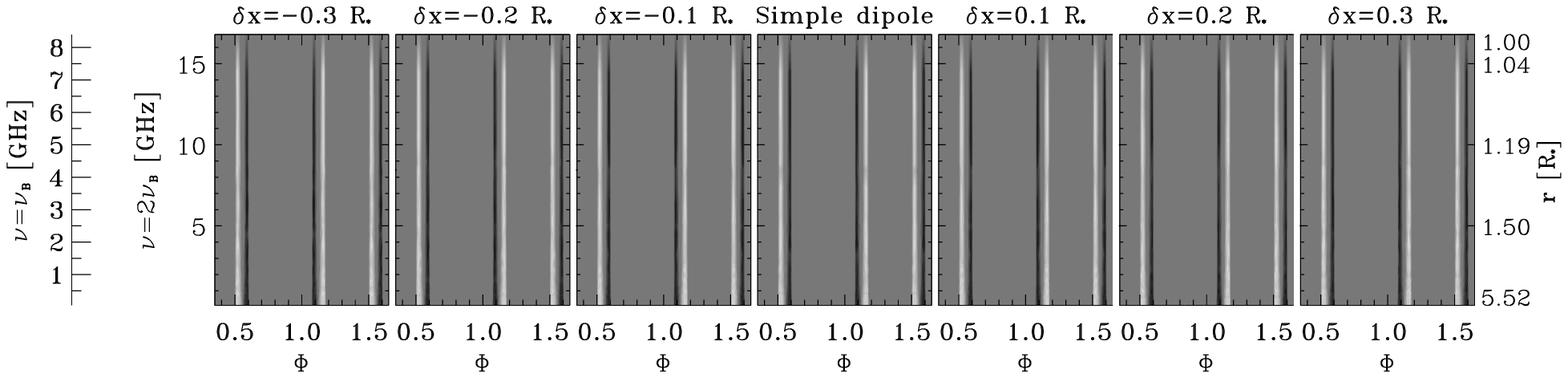}}
\resizebox{\hsize}{!}{\includegraphics{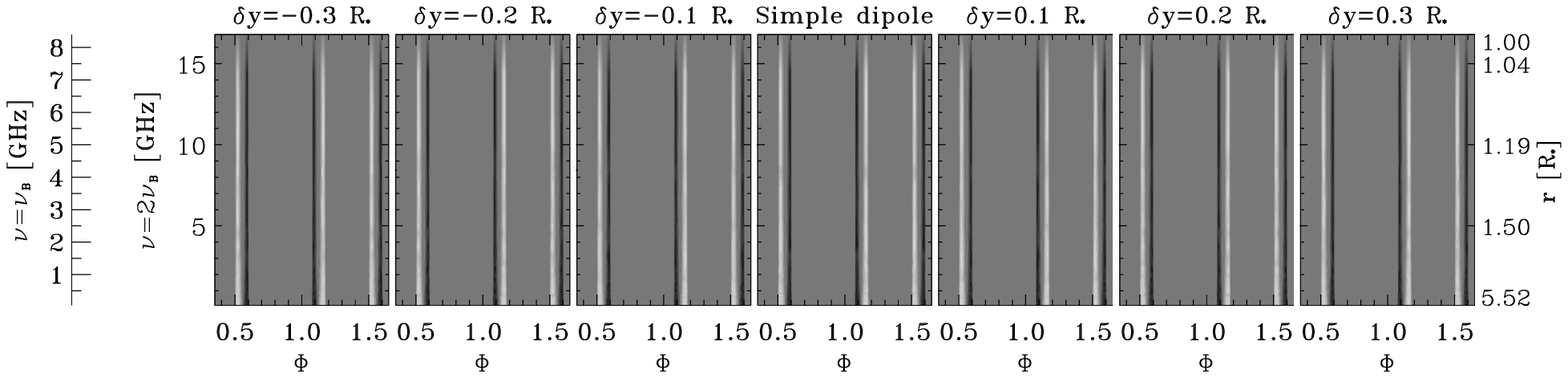}}
\resizebox{\hsize}{!}{\includegraphics{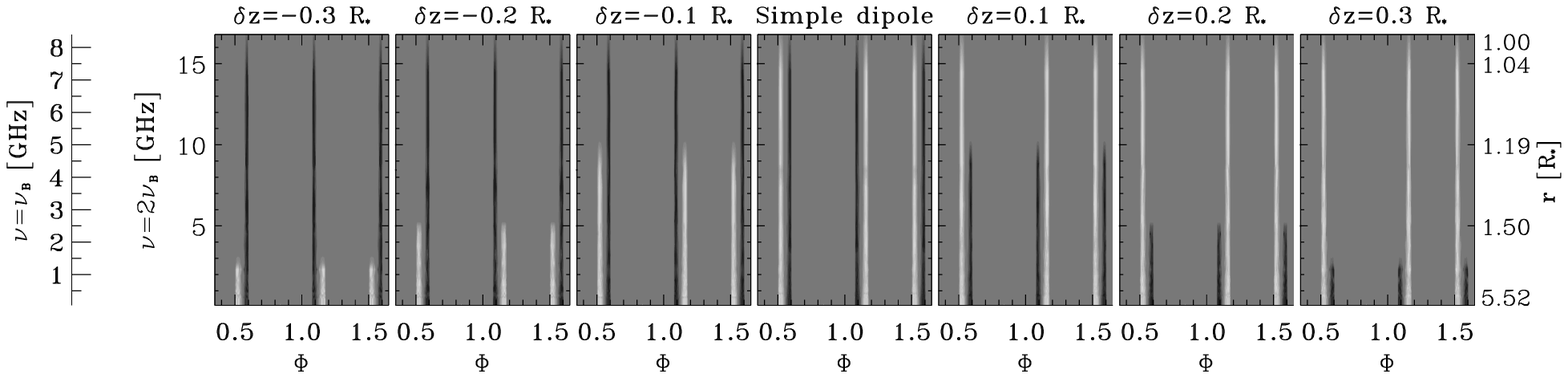}}
\caption{
Simulated dynamic spectra of the auroral radio emission of TVLM\,513-46546
as a function of the dipole shift.
Dipole shifted along the $x$-axis: top panels.  
Dipole shifted along the $y$-axis: middle panels.  
Dipole shifted along the $z$-axis: bottom panels.  
The further axes displayed on the left side of the figures indicates the 
auroral radio emission frequency in case of ECM 
% to the auroral radio emission in case of ECM 
at the first harmonic of the local gyro-frequency.
The corresponding distances of the radiating layers are listed in the right panels axes.
}
\label{test_quadrupolo} 
\end{figure*}
%============================================fig 6

Following these two robust observational constraints, 
we separately analyse the case of a magnetic stellar geometry
with a $B$ maximum at $\Phi \approx 0.85$ (Case 1)
and the case where the stellar geometry is characterised by a $B$ null at $\Phi \approx 0.025$ (Case 2).
The simulated light curves of the auroral radio emission (for the C and X bands), 
that verify the imposed conditions, are shown in Fig.~\ref{tvlm_simulaz}.   
The left panels of Fig.~\ref{tvlm_simulaz} show  the magnetic curves (top panel) and
the modelled auroral light curves, for the X (middle panel) and C (bottom panel) bands.
These curves simulate the auroral pulses from a magnetosphere with an effective magnetic field
curve characterised by extremes falling at  intermediate phases between the C band peaks.
The right panels of Fig.~\ref{tvlm_simulaz} show the simulation from a magnetosphere characterised by an
effective magnetic curve with a null between the doubly-peaked pulses detected in X band.
The magnetic curves that do not invert the sign of the effective magnetic field strength
(absence of nulls) are not shown. This explains the grey bands of the top right panel of Fig.~\ref{tvlm_simulaz}.

In Table~\ref{sol_simul}, we summarise the model solutions achieved separately for the two cases.
For Case 2, the parameter $\Delta \Phi$ is also reported. This parameter is the 
phase shift between the phase location of the magnetic null (negative to positive magnetic field strength) for the two cases analysed.
The parameter can be visually estimated by looking at
the solid lines in the top panels of Fig.~\ref{tvlm_simulaz}.
It also quantifies the different spatial orientation of the magnetospheres 
that satisfies the imposed conditions related to the two geometrical cases under analysis.
The magnitude of the phase shift is necessary to locate exactly the phase position of the 8.44 GHz doubly-peaked auroral pulse and
is a function of the dipole tilt $\beta$.
{We found that the values of this parameter that are able to 
simulate the observed auroral features in both the C and the X bands 
lie in the ranges $100^{\circ}$--$140^{\circ}$ and $220^{\circ}$--$260^{\circ}$,
corresponding to an absolute value of the tilt $\beta$ between the dipole and the rotation axis in the range 
%ranging between $40^{\circ}$ and $80^{\circ}$.
$40^{\circ} < |\beta| <  80^{\circ}$.
A less tilted dipolar magnetosphere is not able to reproduce the auroral pattern observed at 8.44 GHz,
whereas a larger  misalignment of the  oblique rotator cannot reproduce the timing of the pulses detected at 4.88 GHz.

Those simulations performed assuming a simple dipole indicate that 
a phase lag must be introduced between the light curves at the two simulated frequencies,
as evidenced by  $\Delta \Phi \ne 0$. This suggests that}
the orientations of the magnetospheres able to reproduce
the C and the X bands are not the same. %, as evidenced by  $\Delta \Phi \ne 0$.
%does not exist a common magnetospheric orientation
% The orientations of the magnetospheres 
%able to simultaneously reproduce the C and the X bands, as evidenced by  $\Delta \Phi \ne 0$.
{It seems that there is not a common dipolar} geometry able to reproduce simultaneously  
the phase locations of the auroral pulses detected at 4.88 and 8.44 GHz.

{

%________________________________
\section{Auroral radio emission from a non-dipolar magnetosphere}
\label{non_dip}

The model used to simulate the auroral radio emission of TVLM\,513-46546 
was developed under the simplified hypothesis of magnetic field dipolar topology. % of the magnetic field.
But the true magnetic field topology of this fully convective ultra cool dwarf
could be far from a simple dipole.
Recently the magnetic field generation in the low mass rapidly rotating fully convective stars
has been simulated using three-dimensional magnetohydrodynamic models \citep{yadav_etal15}.
These simulations were able to reproduce the kGauss level of the
magnetic field strength measured in some UCDs \citep{reiners_basri07,reiners_basri10}, and
the  dipole-dominated magnetic field topology often observed in the mid M dwarfs
\citep{donati_etal06,morin_etal08a,morin_etal08b}, 
but was also able to simultaneously simulate the presence of a small scale (toroidal) magnetic component,
%at lower magnetic latitudes, 
that characterises some late M type dwarfs \citep{morin_etal10}.

%%============================================fig 6
%\begin{figure*}
%\resizebox{\hsize}{!}{\includegraphics{simul_4pole_b.eps}}
%\caption{\bf cippi cippi bau bau.  
%}
%\label{sim_4pole} 
%\end{figure*}
%%============================================fig 6
%============================================fig 6
\begin{figure*}
\resizebox{\hsize}{!}{\includegraphics{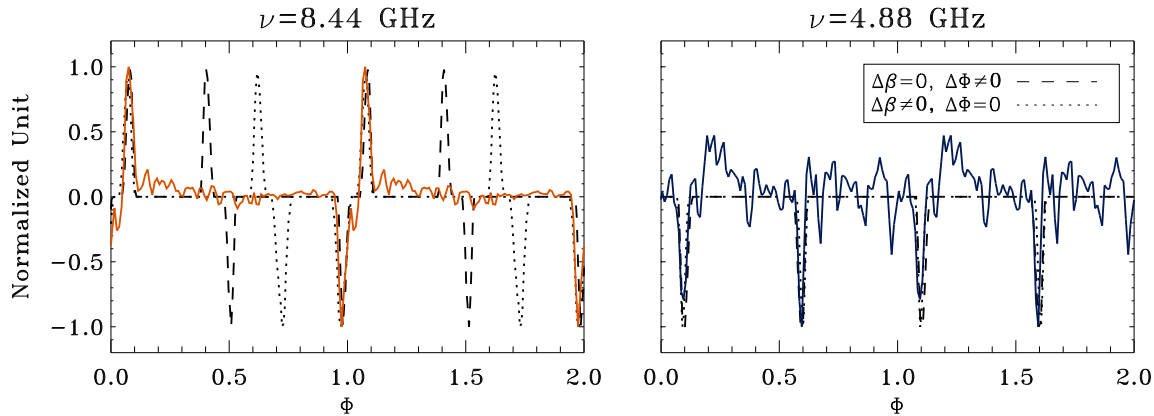}}
\caption{Comparison between the observed and the simulated auroral radio light curves of TVLM\,513-46546. 
Auroral emission at  8.44 GHz (X band): left panel.
Auroral emission at  4.88 GHz (C band): right panel.
%Left panel) 8.44 GHz (X band) auroral emission.
%Right panel) 4.88 GHz (C band) auroral emission.  
The dashed line refers to the simulations performed assuming a common 
magnetospheric geometry for the X and C band emission; to reproduce
the observed timing of the auroral pulses, we introduced a phase-lag between the light curves 
simulated at the two frequencies.
The dotted line refers to simulations arising from magnetospheres with different dipole tilt;
in this case, there is no phase lag between the light curves simulated at the X and C bands.
The C band synthetic curves were simulated assuming a dipole shifted
along the dipole axis ($z$-axis) toward the south pole ($\delta z=-0.3$ R$_{\ast}$).
The X band simulations were performed assuming a simple dipole, 
which is equivalent to assume a dipole shifted along the $x$ or $y$-axes.
}
\label{sim_4pole} 
\end{figure*}
%============================================fig 6

%============================================fig 6

%Citare l'articolo di Berger et al 2009 su 0716 in cui l'autore discute sulla natura non-dipolare del campo
%magnetico di questa altra UCDs.

%Route 2016 discute sulla possibile esistenza di un ciclo magnetico, simile a quello solare che inverte la polarita ogni 11 anni,
%anche sulle UCDs. Route 2016 trae spunto dal fatto che in molte UCDs vengono dettati fenomeni aurorali RCP
%durante certi cicli osservativi, mentre durante altre sessioni vengono
%rivelati impulsi LCP. Questa fatto Route lo interpreta come inversione della polaritˆ magnetic dal campo stellare.

%Io in questo lavoro aggiungo anche la possibilitˆ che questa evidenza osservativa sia la conseguenza di una interazione stella pianeta.

%A) Generic introduction on the magnetic field topology of the rapidly rotating UCDs.
%The magnetic field have a not simple dipolare topology: poloidal plus toroidal component (vedi bib dell tizio che mi ha contattato).

To take into account the non-dipolar topology of the UCD magnetic field,
we modified the model described in Paper I to reproduce the auroral radio emission from a 
stellar magnetosphere shaped like an offset dipole.
This could be assumed as a first approximation
of a non-dipolar magnetic field topology,
and was commonly used to reproduce the complex magnetic field curves measured in some
hot magnetic stars \citep{hatzes97,glagolevskij11,oksala_etal12,oksala_etal15}.

%How to simulate a non dipolar field. A de-centered dipole could be assumed as a rough approximation
%of a non-dipolar magnetic field topology. As example Hatzes~199? using the Doppler Imaging technique 
%describe the non-dipolar field topology of the hot star CU~Vir assuming a de-centered  dipole.

In the following, we analyse the frequency dependence of the stellar auroral radio emission visibility
as a function of the dipole shift, along and across the axis
passing through the centre of the star and parallel to the dipole magnetic axis.
The stellar reference frame ($Oxyz$) is anchored with the stellar centre, has the $z$-axis
parallel or coincident with the dipole axis, and the $x$-axis is located in the plane passing through
the $z$-axis and containing the rotation axis.
%in the reference $Oxyz$ anchored with the star, where the $z$-axis coincides with the dipole magnetic axis.
Details of the method used
%enabled to simulate the auroral radio emission from an offset dipole 
are given in Appendix~\ref{offset_dip}. 

In Fig.~\ref{test_quadrupolo}, we show the simulated dynamical spectrum
of the auroral radio emission from the magnetosphere of TVLM\,513-46546 shaped like an offset dipole,
and constrained to have the maximum magnetic field strength at the stellar surface not exceeding 3000 Gauss.
The frequency range analysed is 16.88--0.1 GHz
(8.44--0.05 GHz if the auroral emission is generated at the first harmonic of the local gyro-frequency).
The higher frequency limit has been chosen on the basis of the value of the
first harmonic of the gyro-frequency at the stellar surface ($B_{\mathrm p}=3000$ Gauss), whereas the low-frequency limit is 
close to the lower limit of the frequency range of the new-generation Earth-based low-frequency radio interferometers,
such as the Low-Frequency Array (LOFAR; \citealp{van_haarlem_etal13}), the Murchison Widefield Array (MWA; \citealp{tingay_etal13}), 
or the forthcoming Square Kilometre Array (SKA).
%\textcolor{cyan}{Need references regarding LOFAR, MWA and SKA?}.
The selected frequency range 
corresponds to the heights above the stellar surface that lie in the range $\approx0$--4.5 [R$_{\ast}$]. %0--2.23 [R$_{\ast}$].
% in case of auroral emission at the fundamental gyro-frequency, or ??--?? in case of emission at the second harmonic.
The simulations have been performed assuming for each analysed frequency the same emission beam pattern
(in particular $\delta=15^{\circ}$ and $\theta=10^{\circ}$) and a common tilt of the magnetic axis ($|\beta|=60^{\circ}$).
The simulations have been performed varying the dipole shift along the $xyz$-axes
in the range from $-0.3$ to 0.3 [R$_{\ast}$], with a step of 0.1 [R$_{\ast}$].

The results of the dynamical spectra simulations from an offset-dipole are shown in Fig.~\ref{test_quadrupolo}.
By looking at the figure, it is clear that
the dipole offset along the $x$-axis or the $y$-axis has negligible effect on the simulated dynamic spectra
and consequently on the auroral light-curve shape.
In these cases, 
%the simulated effective magnetic curves evidence a clear enhancement of their complexity
%increasing the dipole shift from the center of the star (see the top panels of  Fig.~\ref{test_quadrupolo} referred to the 
%cases of shift along the $x$ and $y$-axes), 
%but the phase location of the magnetic nulls are always unchanged, and
the corresponding simulated dynamic spectra are almost indistinguishable from the
case of auroral emission from a simply dipolarly-shaped magnetosphere. 
%
%%When the dipole moves along the dipole axis
%When the dipole is shifted along the $z$-axis, coinciding with the dipole axis, 
%%the phase locations of the magnetic nulls will be variable, and 
%above a limit value of $\delta z$ 
%the stellar effective magnetic curve does not change its polarity.
%%The case of dipole shiftted along the dipole axis ($z$-axis) 
%
%this case determines a north-south asymmetry
%
When the dipole is shifted along the $z$-axis, coinciding with the dipole axis, 
this determines a north-south asymmetry
that significantly affects the stellar auroral radio-emission visibility
(see bottom panels of  Fig.~\ref{test_quadrupolo}). % referring to the shift along the $z$-axis).
%The auroral signatures from each stellar hemispheres become
%progressively asymmetric as the dipole shift enhances.
The spectral dependence of the auroral radio emission arising from
each stellar hemisphere becomes
progressively asymmetric as the dipole shift increases.
The auroral radio emission arising from the hemisphere
where the dipole approaches the stellar surface is almost unchanged,
whereas the emission from the opposite hemisphere progressively disappears, as the radio frequency increases.
This effect depends on the amplitude of the dipole shift from the centre of the star.

%To generalize the results of the above simulations,
%in the Appendix~\ref{size_depend} the dependence of the auroral radio emission from the parameters
%that control the emission beam pattern and the magnetospheric size is explored.
%These new simulations confirm the results presented in Paper I, 
%obtained for the case of the auroral radio emission arising from a simple dipolar magnetosphere.

To deeply explore the effect of the offset dipole on the auroral radio emission visibility
from TVLM\,513-46546, % at the C and X observing bands,
%we extended the sets of free parameters that control the auroral beam pattern and the stellar geometry, %that control the auroral beam pattern 
we performed simulations, tuned at the C and X observing bands,
using the parameters listed in Table~\ref{sol_simul}.
The dipole location was shifted along the $xyz$-axes.
%and varying the dipole shift along the $xyz$-axes.
%we performed new simulations %simulated the auroral radio emission of TVLM\,513-46546
%as a function of the dipole shift along the $xyz$-axes, and
%%varying the amplitude of the dipole shift along the $xyz$-axes, and
%using the parameters that control the auroral beam size and the stellar geometry listed in Table~\ref{sol_simul}.
These new sets of simulations are pictured in Figs.~\ref{case1_4pole} and \ref{case2_4pole}. % Appendix~\ref{app}. % and shown in Figs.~\ref{case1_4pole} and \ref{case2_4pole}.

As examples of the simulated auroral radio light curves, we show in Fig.~\ref{sim_4pole} the observed auroral emission with  
two representative simulated curves superimposed; the left panel refers to the pulses detected at 8.44 GHz, the right panel to that at 4.88 GHz.
The two sets of model parameters used to simulate the auroral light curves shown in Fig.~\ref{sim_4pole}
were selected to reproduce two possible cases. In the first case, 
%there is not phase lag between
we simulated 
%the simulated 
light curves at 4.88 and 8.44 GHz without phase lag, but the two corresponding stellar geometries
are not the same ($\Delta \Phi=0$ and $\Delta \beta \neq 0$):
the adopted dipole tilts %for the case of non-shifted light curves 
are $\beta=40^{\circ}$ at 8.44 GHz and $\beta=100^{\circ}$ at 4.88 GHz
($\Delta \beta$ minimun).
In the second case, we simulated the auroral emission at the two observing frequencies
using the same stellar geometry (in particular we assumed  $\beta =120^{\circ}$), 
but we needed to introduce a phase lag between the two light curves at the two frequencies
($\Delta \beta = 0$ and $\Delta \Phi \neq 0$).
%\textcolor{red}{rifrasare la frame in italico, non si capisce}{\it The two simulations shown in Fig.~\ref{sim_4pole} have been performed 
%assuming that the auroral radio light curves simulated at 4.88 and 8.44 GHz are not shifted ($\Delta \Phi=0$ and $\Delta \beta \neq 0$),
%and assuming the same magnetic geometry but shifting the light curves at the two frequency (in this case $\beta =120^{\circ}$).}
%the adopted dipole tilts %for the case of non-shifted light curves 
%are $\beta=40^{\circ}$ at 8.44 GHz, and $\beta=100^{\circ}$ at 4.88 GHz
%($\Delta \beta$ minimun).
%The chosen value of the dipole tilt reproducing the timing of the pulses detected at 4.88 GHz, minimises the difference $\Delta \beta$
%with the unique allowed magnetospheric geometry that simulates a non-shifted 8.44 GHz auroral light curve.
The corresponding combination of the parameters that control the auroral beam pattern are listed in Table~\ref{sol_simul}.

%\textcolor{red}{rifrasare e spezzare la frame in italics, troppo lunga e poco chiara}{\it Even if also assuming a simply dipolar shaped magnetosphere exist sets of model parameters that are able to
%simulate C band auroral emission with only two  left-hand circularly polarised pulses per stellar rotation (see Table~\ref{sol_simul}), 
%like in the observations,
%%The inspection of the solutions {\it derived assuming a simply dipolar shaped magnetosphere}, reported in Table~\ref{sol_simul},
%%highlights that certain sets of model parameters are able to
%%simulate C band auroral emission with only two  left-hand circularly polarised pulses per stellar rotation, like in the observations.
%%Furthermore, 
%the synthetic curves at 4.88 GHz, pictured in Fig.~\ref{sim_4pole}, was obtained in the case of a dipole shifted along the
%dipole axis (in particular $\delta z=-0.3$ R$_{\ast}$), 
%whereas the curves at 8.44 GHz was obtained assuming a simple dipole.}

The 4.88 GHz synthetic curves in Fig.~\ref{sim_4pole} were obtained using a dipole shifted along the
dipole axis (in particular $\delta z=-0.3$ R$_{\ast}$). The curves at 8.44 GHz was instead obtained assuming a simple dipole,
which is equivalent to the cases of an offset-dipole along the $x$ or $y$-axes.
The effect of the dipole shift along the $z$-axis makes it easier to reproduce the auroral peaks
detected at the lower frequency (4.88 GHz). In fact, the auroral component arising from the northern
hemisphere, expected RCP circularly polarised, is naturally suppressed by the displacement
of the dipole in the direction to the south pole.
The auroral features detected at the higher frequencies (8.44 GHz) are instead not compatible with
an offset dipole along the polar axis. 
In fact, a general result of our simulations is that if the northern signature of the auroral emission at 4.88 GHz disappears,
then there cannot be auroral radio emission radiated at frequencies higher than 4.88 GHz.

%{\it 
%The need to use simulated auroral light curves at 4.88 and 8.44 GHz
%obtained using magnetospheres with different dipole tilt,
%or, if fixed the dipole tilt, shifted in phase, %$\Delta \Phi \neq 0$,
%indicates that also in the case of the auroral radio emission from an offset dipole
%we cannot find a common magnetospheric geometry that is able to simultaneously reproduce
%the timing of the auroral pulses detected at the two observing radio frequencies.
%}
%The need to use simulated auroral light curves at 4.88 and 8.44 GHz
%obtained using magnetospheres with different dipole tilt,
%or, if fixed the dipole tilt, shifted in phase, %$\Delta \Phi \neq 0$,
%indicates that also 
In the case of the auroral radio emission from an offset dipole,
we have not found a common magnetospheric geometry that is able to simultaneously reproduce
the timing and pattern of the auroral pulses detected at the two observing radio frequencies (4.88 and 8.44 GHz).
%and the lack of right hand circularly polarised (Stokes ${\mathrm V}>0$) C band auroral emission.
We reach  the same conclusion obtained for the case of the auroral radio emission arising from a simple dipolar magnetosphere (Sec.~\ref{model}).
The geometry of the TVLM\,513-46546 magnetosphere able to reproduce the auroral pulses detected at 8.44 GHz is not the same as that adopted to simulate the 4.88 GHz pulses.

}

{
%________________________________
\section{The magnetosphere of TVLM\,513-46546}
\label{mag_tv}

%The detection of the auroral radio emission

%The detection of the auroral radio emission is strongly dependent on the magnetic field vector orientation with respect to the line of sight,
%which is related to the topology of the magnetic field in the region where it took place.
%This coherent phenomenon can be used to achieve indirect informations regarding the geometry of the stellar magnetosphere.
%In fact, to reproduce the features of the auroral radio emission of TVLM\,513-46546, detected at 4.88 (C band) and 8.44 (X band) GHz,
%%the simulations performed in this paper 
%we constrained the inclination of the stellar magnetic axis.
%%that is able to reproduce the features of the auroral radio emission of TVLM\,513-46546 detected at 4.88 (C band) and 8.44 (X band) GHz.
%On the other hand, the simulations, performed assuming both a simple dipole (Sec.~\ref{model}), and an offset dipole (Sec.~\ref{non_dip}), 
%highlight that there is not a common geometry able to reproduce simultaneously 
%the phase locations of the auroral pulses detected at 4.88 and 8.44 GHz.

%\subsection{Time stable magnetic field}
%\label{time_stable}

Like the case of Jupiter, where the study of the planet
auroral radio emission was able to constrain the topology of its magnetosphere
\citep{hess_etal10,hess_etal11}, % ({\it controlla references}),
%the analysis of the auroral radio emission of the UCD TVLM\,513-46546 could be used 
%as a tool to probe the magnetospheric regions where the pulses detected at 4.88 and 8.44 GHz originate.
the analysis of the stellar auroral radio emission arising from the UCD TVLM\,513-46546 
can be used to obtain indirect information regarding the overall field topology of the stellar magnetosphere.
%can be useful to obtain indirect information regarding the overall field topology of its magnetosphere.
%can be used to achieve indirect informations regarding the geometry of its magnetosphere.
%In fact, to reproduce the features of the auroral radio emission of TVLM\,513-46546, detected at 4.88 (C band) and 8.44 (X band) GHz,
%%the simulations performed in this paper 
%we constrained the inclination of the stellar magnetic axis.
%In general, 
The %wide band 
auroral radio emission gives a snapshot of the stellar magnetic field topology
at the epoch of its detection,
and its modelling has been already used as a method to describe the field topology of UCD magnetospheres \citep{metodieva_etal17}. 

%This coherent phenomenon can be used to achieve indirect informations regarding the geometry of the stellar magnetosphere.
To use this coherent phenomenon %the stellar auroral radio emission
%, collected on different epochs, 
as a tool to investigate
the stellar magnetospheric configuration, it is necessary to assume the temporal stability of the stellar magnetic-field topology
over the time range spanned by the measurements. 
%
%The here analysed auroral radio emission from TVLM\,513-46546 was measured
%during two consecutive days of May 2006, each single observing run performed
%at the same frequency, respectively in the C and X band \citep{hallinan_etal07}.
%
%This analysis has been performed under the assumption of the temporal
%stability of the magnetic structures where the auroral radio emission take place. 
In the case of TVLM\,513-46546,}
even though the two bands are not observed simultaneously,
this is a fully sensible hypothesis, also supported 
by the almost simultaneous C and X broad-band observations performed with the
Karl G. Jansky Very Large Array (JVLA)
during other active epochs of TVLM\,513-46546, 
which display a very similar behaviour \citep{lynch_etal15}.
{
This similarity supports the hypothesis of the temporal stability of the overall magnetic-field topology
of the TVLM\,513-46546 magnetosphere, over a time  of $\approx 10$ years.

%\subsection{Torqued magnetic field line}
\subsection{Torqued magnetosphere}
\label{bent_line}

%The simulations, performed assuming both a simple dipole (Sec.~\ref{model}), and an offset dipole (Sec.~\ref{non_dip}), 
%highlight that there is not a common geometry able to reproduce simultaneously 
%the phase locations of the auroral pulses detected at 4.88 and 8.44 GHz.

The modelling of the phase occurrence of the TVLM\,513-46546 pulses (Sections~\ref{model} and \ref{non_dip})
%detected respectively at 4.88 and 8.44 GHz 
highlights that %, to reconcile the simulations with the observations, 
the beams of the
auroral emission at 8.44 (X band) and 4.88 GHz (C band) have to be misaligned.
As discussed in Paper I,
the elementary physical process responsible for the stellar auroral radio emission
is the Electron Cyclotron Maser (ECM). The orientation of the highly-beamed ECM emission pattern
%is strongly related to the orientation of the local magnetic field vector.
is strongly related to the orientation of the magnetic-field vector in the region where it takes place.
%The auroral radio emission is strongly dependent on the magnetic field vector orientation with respect to the line of sight,
%which is related to the topology of the magnetic field in the region where it took place.
%
The phase lag of the 4.88 GHz pulses could then be explained as a consequence of
a different orientation of the magnetic-field vector
%longitudinal location 
with respect to the auroral region where the
8.44 GHz emission takes place.
%
%The 
%%To well display the 
%geometry of the stellar auroral emission is displayed
%in Fig.~\ref{fig_bent}, where the orientations of the auroral beams
%at the two radio frequencies here analysed are shown.
%In figure it is pictured the case related to the minimum value of $\Delta \Phi$
%listed in Table~\ref{sol_simul} ($\Delta \Phi=0.08$ with $\beta=100^{\circ}$).
%
This
%To well display the 
%geometry %of the stellar auroral emission 
is displayed
in Fig.~\ref{fig_bent}, where the orientations of the auroral beams
at the two radio frequencies here analysed are shown.
The  figure is related to the minimum value of $\Delta \Phi$
listed in Table~\ref{sol_simul} ($\Delta \Phi=0.08$ with $\beta=100^{\circ}$).
It is a consequence that the magnetic field lines passing trough the cavity boundaries where the two frequencies arise have to be bent.
%%As previously discussed,
%%the elementary physical process responsible for the stellar auroral radio emission
%%is the Electron Cyclotron Maser, and
%%the orientation of the highly beamed ECM emission pattern
%%is strongly related to the orientation of the local magnetic field vector.
%In principle, if the pulses at 4.88 and 8.44 GHz originate from the same auroral cavity,
%the corresponding magnetic field vectors in the layers where the two frequencies originates
%%that are located in the meridian plane containing the magnetic field lines passing trough 
%%the boundary of the magnetic cavity 
%have the same orientations with respect to the line of sight.
%Follow that there is any phase lag between the auroral pulses radiated at the two frequencies.
%If the pulses at 4.88 and 8.44 GHz originate from
%the same magnetic cavity, then their %boundary 
%magnetic field lines have to be bent.
The dynamics of the plasma of the fast rotating UCD TVLM\,513-46546  could give rise to a torqued magnetosphere,
like in  Jupiter \citep*{hill_etal74,hill_79}.
In principle, the auroral radio emission arising from a torqued magnetosphere 
could explain the timing of the radio pulses from TVLM\,513-46546,
but this scenario is affected by a paradox.

%============================================fig 5
\begin{figure}
\resizebox{\hsize}{!}{\includegraphics{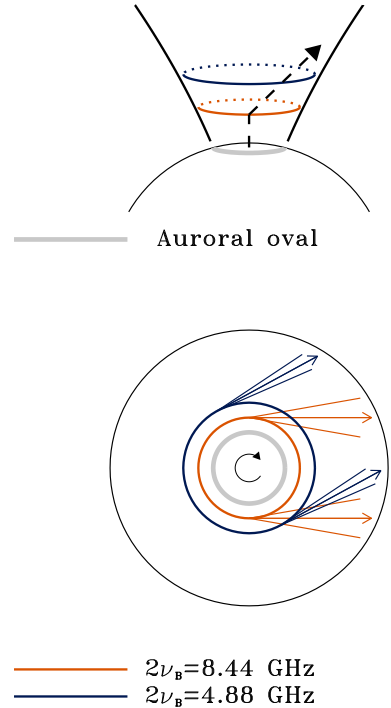}}
\caption{Top panel, side view of the auroral cavity 
that radiates at 4.88 and 8.44 GHz (assuming $B_{\mathrm p}=3000$ [Gauss]).
The footprint at the stellar surface is also pictured (auroral oval).
In the bottom panel is shown the top view of the auroral cavity. 
The auroral beam pattern at the two analysed frequencies are shown.
The corresponding bended magnetic field line is schematically pictured in the top panel (dashed line).
}
\label{fig_bent}
\end{figure}
%%============================================fig 5

%In fact, 
In the case of a simple dipolar magnetic topology (Sec.~\ref{model}), 
our simulations evidence that there is no stellar geometry 
able to produce LCP C band auroral radio emission
%that is detectable from the souther hemisphere only,
and simultaneously LCP+RCP X band pulses.
To reproduce auroral radio emission arising at a given radio frequency
from only one hemisphere, one needs to assume an offset dipole along the 
polar axis ($z$-axis), see Sec.~\ref{non_dip}.
Due to the direct stellar magnetic-field strength dependence, 
the X band (8.44 GHz) pulses originate from a magnetospheric region closer
to the stellar surface, when compared with the location of the
ring in which the C band (4.88 GHz) pulses arise.
In the top panel of Fig.~\ref{fig_bent}, we show the side view of the auroral rings 
above the magnetic pole where the pulsed emission of TVLM\,513-46546 originated.
The simulations with an offset dipole show 
that
the auroral component arising from the hemisphere
from which the dipole is moving away
suffers from a high-frequency cutoff, as the dipole is shifted along the polar axis
(see bottom panels of Fig.~\ref{test_quadrupolo}). % referred to the case of a dipole shift along the $z$-axis).
The auroral emission at the radio frequencies generated close to
the star are affected earlier by the offset dipole.

%On the countary,
The paradox is that while
the auroral emission at 8.44 GHz
arising from the northern hemisphere should be suppressed, the RCP component (Stokes $\mathrm{V} >0$) of the X band auroral pulse is  instead clearly detected (left panel of Fig.~\ref{sim_4pole}).

\subsection{Two distinct auroral cavities}
\label{multi_aurora}

As previously discussed (Sec.~\ref{non_dip}), the magnetic-field topology of
 fast-rotating fully-convective low-mass stars could be characterised by the coexistence of large- and small-scale magnetic fields \citep{yadav_etal15}.
The large-scale field component is well described by a magnetic field with an overall axial symmetry (dipole or offset dipole),
whereas the small-scale magnetic-field components have a toroidal topology.
For the fast-rotating UCD TVLM\,513-46546, it is therefore reasonable to assume a similar magnetic-field configuration.

%Then, the paradox raised in the previous section
%This paradox 
The difference between the C and X band TVLM\,513-46546 auroral emission
%measured on TVLM\,513-46546,
%The auroral patterns measured 
could be explained if
the coherent pulses at 4.88 and 8.44 GHz originate from two different auroral cavities,
related to two distinct magnetic-field components. 
The cavity where the LCP pulses at 4.88 GHz originate would be related to a large magnetic cavity with footprints close to the poles,
shaped like a dipole shifted along the polar axis.
%with the centrifugally opened field lines filled by ionised material, 
%that is not rigidly corotating with the star  \citep{nichols_etal12}. % like the  \citep{hill_etla74,hill_79}.
%The cavity where the 4.88 GHz originates is associated 
%with the centrifugally opened field lines filled by ionised material, 
%that is not rigidly corotating with the star  \citep{nichols_etal12}. % like the  \citep{hill_etla74,hill_79}.
The LCP+RCP 8.44 GHz auroral pulses would instead arise from a more internal cavity, 
related to small-scale magnetic structures (like magnetic loops with equator-ward footprints).
If the regions with opposite magnetic-field polarity 
of this small-scale auroral cavity
both have magnetic-field strength large enough
to radiate auroral radio emission at 8.44 GHz, then
auroral pulse components of opposite polarisations will be detectable.
%By following a different simulation approach, \citet{lynch_etal15} proposed a similar scenario.
%Whereas the 8.44 GHz auroral pulse originates from an innermost cavity 
%related to the toroidal magnetic field component  in a neutral medium.
The meridian plane containing the magnetic-field lines delimiting the
small-scale auroral cavity may not be closely related with
the geometry of the large-scale magnetic-field topology.
%to the overal stellar magnetospheric geometry.
The large- and small-scale auroral cavities could then
be described by two distinct magnetic-dipole orientations.
The possible existence of two distinct auroral cavities 
could therefore explain the phase lag of the 4.88 GHz pulses with respect to those detected at 8.44 GHz. %the higher frequency.

%Then, these two possible auroral cavities could not have the same orientation,
%explaining the phase lag of the 4.88 GHz pulses with respect to those detected at the higher frequency.
%respect those detected at the higher frequency.
%the two auroral cavities have not the same orientation.
%that explaining the lag go the 4.88 GHz pulses respect those detected at the higher frequency.

%Within this scenario, the stellar auroral radio emission of  TVLM\,513-46546 could be useful
%to quantify the magnetic field strength at the footprints of the equator ward field lines.

%Assuming that the auroral radio emission from two different cavities does not contaminate each other,
The non-detection in the C band of the auroral features observed in the X band (LCP+RCP auroral pulse),
%can be used to constrain the magnetic field strength at the footprints of the equator ward field lines.
%
%The non-detection at the X band of the auroral features observed at the C band, and viceversa,
%can be used to constrain the magnetic field strength of the two
%cavities, where each component of the auroral radio emission originates.
and viceversa,
can be used to constrain the magnetic-field strength
of the footprints of the auroral cavities.}
In the case of the loss-cone driven ECM emission mechanism,
the deflection angle $\theta$ is directly related to the hollow cone half-aperture $\theta_{\mathrm B}$,
where the radiation amplified by the elementary emission process is beamed
(in the case of the northern hemisphere auroral emission $\theta=90^{\circ}-\theta_{\mathrm B}$).
Furthermore, the hollow cone half-aperture is a function of the frequency of the amplified
radiation as follows:
$\cos \theta_{\mathrm B}  = v/c \sqrt{1-\nu_{\mathrm B}/\nu_{\mathrm {B_{max}}}}$  \citep{hess_etal08},
with $v$ the speed of the unstable electrons and $\nu_{\mathrm {B_{max}}}$ gyro-frequency at the stellar surface.
The model free parameters ($\theta$ and $\delta$)
that reproduce the auroral radio emission detected at 8.44 GHz are reported in Table~\ref{sol_simul}.
In the cases of the solutions for the X band, if the emission beam pattern is deflected of further 2.5--5 degrees 
it does not intercept the line of sight and the auroral radio emission will not be detectable.
Assuming that the 4.88 GHz emission suffers from the above determined further deflection,
it is then possible to estimate the gyro-frequency at the stellar surface and then the magnetic field strength
by using the relation between $\theta_{\mathrm B}$ and $\nu_{\mathrm {B_{max}}}$ written above,
given the deflection angles at two different values of the gyro-frequency.
The corresponding values of the magnetic-field strength at the stellar surface lie in the range $\approx 2000$--3300 Gauss.
{The estimation above gives only a rough constraint to the magnetic-field strength.
%The measurement of 
To obtain a more refined magnetic-field strength estimation
at the footprints %of opposite magnetic polarity 
of the equator-ward toroidal magnetic-field lines, 
it is necessary to measure the high-frequency cutoff of the auroral pulses of
opposite polarisation. This measurement can be obtained only by acquiring the wide-band dynamical spectrum 
of the stellar auroral radio emission.

At the opposite of the 8.44 GHz emission, the}
%If the amplified emission frequency is the first harmonic of the local gyro-frequency, then
%the estimated values of the magnetic field strengths double.
%The 
4.88 GHz coherent pulses take place
in the large auroral cavity ($L=15$--40 R$_{\ast}$ corresponding to $75^{\circ} < \lambda < 81^{\circ}$)
{and do not have a counterpart at 8.44 GHz (Fig.~\ref{tvlm_dic_spec} and Fig.~\ref{sim_4pole}).}
%and are not detected at 8.44 GHz (Fig.~\ref{tvlm_dic_spec}).
This high frequency cutoff can be easily explained if the magnetic field strength {at the south pole
is $< 1500$ Gauss}. %, describing the larger magnetospheric regions with poleward open field lines.
In this case, the second harmonic at the stellar surface is $< 8.44$ GHz
and the corresponding ECM emission will be not amplified.
If the amplified emission frequency is the first harmonic of the local gyro-frequency, then
the estimated values of the magnetic field strengths double.

%============================================fig 8
\begin{figure}
\resizebox{\hsize}{!}{\includegraphics{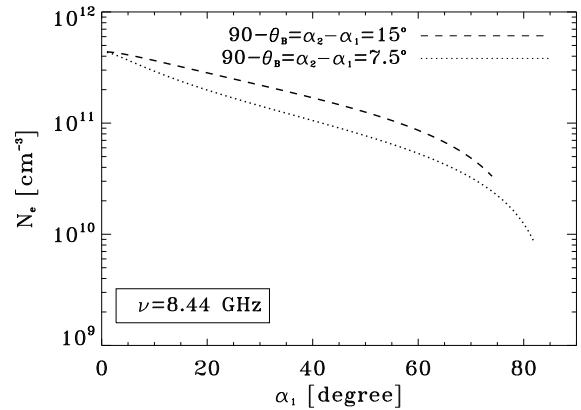}}
\caption{Density of the refractive plasma as a function of the incident angle 
able to deflect the X band {($\nu=8.44$ GHz)} auroral emission,
amplified by a pure horseshoe driven ECM mechanism, of the values that fit the observations.
}
\label{plot_refr}
\end{figure}
%============================================fig 8

The discussion made so far has been done under the assumption of vacuum propagation.
The possible existence of layers of thermal plasma around the star implies the deflection of the ray path.
This additional contribution to the ray path deflection needs a lower hollow-cone
half aperture (for the loss-cone driven ECM emission mechanism)
{
to make the auroral radio emission of the innermost cavity undetectable at the lower frequency ($\nu \le 8.44$ GHz),
thus causing the enhancement of the magnetic-field strength at the stellar surface.
}

The refractive effects are quantified by the Snell~law: $\sin \alpha_{\mathrm r}= \sin \alpha_{\mathrm i} /n_{\mathrm {refr}}$,
where $\alpha_{\mathrm i}$ and $\alpha_{\mathrm r}$ are the angles of incidence and refraction and $n_{\mathrm {refr}}$ 
the refractive index of the medium.
In the case of the extraordinary magneto-ionic mode and for a propagation parallel to the local magnetic field, 
the refractive index can be written as $n_{\mathrm {refr}}^2=1-\nu_{\mathrm {p}}^2 / (\nu^2 - \nu \nu_{\mathrm B})$
($\nu_{\mathrm {p}}\approx 9 \times 10^{-6} \sqrt{N_{\mathrm e}}$ GHz 
plasma frequency of the thermal electrons with number density $N_{\mathrm e}$).
If the auroral radio emission is amplified by a pure horseshoe-driven ECM elementary emitting process,
the ray path is perpendicular to the local magnetic field at all frequencies.
Under this assumption, it is possible to evaluate the
electron number density necessary to deflect
the horseshoe-driven auroral emission exactly of the angle $\theta$, reported in Table~\ref{sol_simul}.
In Fig.~\ref{plot_refr} we show the electron number density of the refractive layer,
that deflects the auroral ray path generated at 8.44 GHz, as a function of the incident angle.
The values of electron density thus estimated are incompatible with the wave propagation at 4.88 GHz.
In fact,  the lower frequency waves cannot freely propagate at the corresponding refraction indices.
This allows us to exclude the horseshoe-driven ECM as the leading
emission mechanism, confirming the conclusion of \citet{lynch_etal15}
that the auroral radio emission of TVLM\,513-4654 is  driven by the loss-cone instability.
However, the presence of a lower density refractive plasma ($N_{\mathrm e} < 10^{10}$ cm$^{-3}$) 
that concurs to the ray path defection of the auroral plasma cannot be ruled out.
Furthermore, the spatial orientation of such refraction planes can introduce 
a frequency-dependent longitudinal deflection of the auroral radiation.
This effect can take into account the measured phase shift between the light curves at 
4.88 and 8.44 GHz.

%============================================fig 5
\begin{figure}
\resizebox{\hsize}{!}{\includegraphics{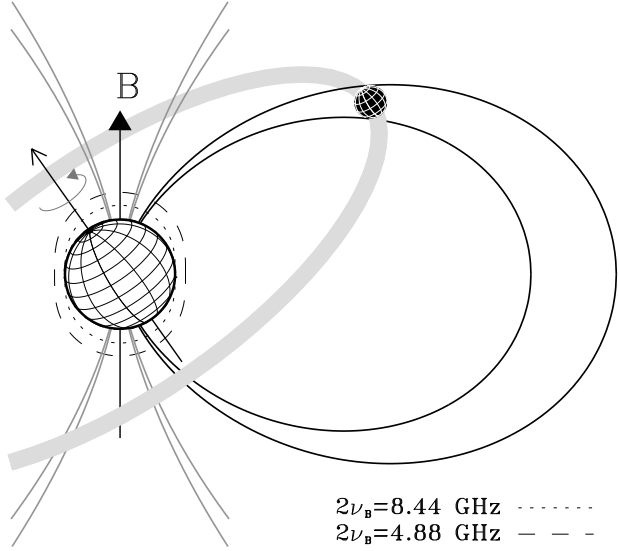}}
\caption{Planet orbit within the magnetosphere of TVLM\,513-46546; angles and size are not in scale.
The magnetic field lines that are crossed by the planet are displayed using the thick black lines. 
The magnetic field lines of the cavity with footprints at high magnetic latitudes
are shown by the light grey magnetic field lines.
The cross section of the equipotential surface of a simple dipole with $B_{\mathrm P}=3000$ Gauss
is also displayed. The magnetic strengths analysed
correspond to the second harmonic of the local gyro-frequency that are equal to 4.88 and 8.44 GHz.  
}
\label{sezione}
\end{figure}
%============================================fig 5

%________________________________
\section{Planet-induced auroral emission}
\label{planet}

%{\it The cavity where the 4.88 GHz arises is associated with the centrifugally opened field lines filled by ionised material, 
%that is not rigidly corotating with the star  \citep{nichols_etal12}. % like the  \citep{hill_etla74,hill_79}.
%}

{To explain the observed differences between the auroral pulses detected at 4.88 and 8.44 GHz (the
phase lag and the circularly polarised features),
%To explain the observed different phase lag and circularly polarised features between the auroral pulses detected
%at 4.88 and 8.44 GHz, % from TVLM\,513-46546, 
we have examined two possible scenarios in the previous section.
In particular, we propose the existence of two distinct auroral cavities,
%on the magnetosphere of TVLM\,513-46546, 
where the emission at the two observing frequencies originates.}

The analysis performed in the {Sec.~\ref{model} and Sec.~\ref{non_dip}} highlights how the auroral radio emission
arising from the whole stellar magnetosphere generates two pulses per stellar
rotation, each pulse being singly or doubly peaked with opposite polarisation sense.
On the contrary, the auroral features of TVLM\,513-46546 detected at 8.44 GHz
are characterised by only one doubly-peaked pulse per stellar rotation.
{Such observational evidence could be explained if the auroral cavity, 
where the 8.44 GHz emission originates, is constrained within a small range of magnetic longitudes.

%like the magnetic arcades surrounding the Sun active regions.
%But, in the case of the UCD TVLM\,513-46546,
%these active magnetic structures have to be stable over a time scale of years.
To explain some features of the auroral radio emission of  UCDs,
%The auroral radio emission from the UCDs is also explained following 
%To reproduce the timing and the polarised features of the X band auroral light curve,
%In the following
%We analyse if 
%we have 
some authors have proposed a scenario like the Io-Jupiter interaction, where the Io-DAM emission originates in the Io's flux tube. Recently,}
%could be plausible in the case of TVLM\,513-46546.}
%we take into account a scenario like the Io-Jupiter interaction, where the Io-DAM emission originates in Io's flux tube.
\citet{hallinan_etal15} suggested  the existence of magnetic interaction between the stellar magnetosphere
and one orbiting exoplanet to explain the auroral radio emission modulation of the UCD LSR\,J1835+3259.
This supports the idea that, at the bottom of the main sequence, the 
stellar auroral radio emission can be also powered by the magnetic interactions  
with rocky planets orbiting around ultra cool and brown dwarfs. %, as suggested by  \citet{hallinan_etal15}.
The indication of the existence of big dust grains (millimetre size) in debris disks around brown dwarfs
\citep{ricci_etal12,ricci_etal13} makes it possible to form rocky planets around very low-mass UCD stars.
{Direct confirmation of the existence of rocky planets around the UCDs is given by the
recent discovery of Earth like planets orbiting around the M8 type dwarf 2MASS\,J23062928-0502285 (TRAPPIST-1) 
\citep{gillon_etal17}.}

%============================================fig 6
\begin{figure*}
\resizebox{\hsize}{!}{\includegraphics{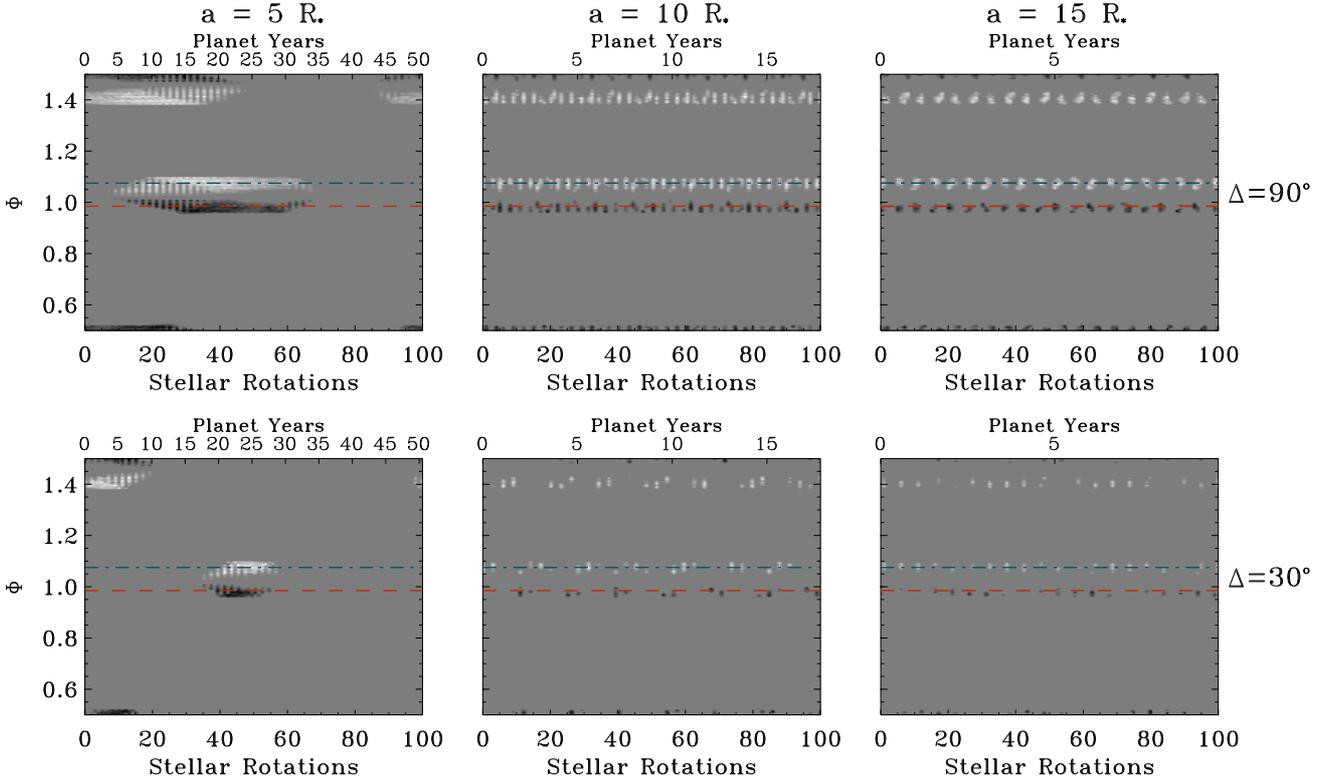}}
%\resizebox{\hsize}{!}{\includegraphics{ecm_planet_plan_orbit30deg.eps}}
\caption{
Simulated evolution of the auroral radio light curve of TVLM\,513-46546.
{The planet orbital plane was assumed coinciding with the stellar rotation plane.}
The light curves are calculated over a time range that spans 100 stellar rotations.
The displayed dynamic light curves simulate the
auroral radio emission arising from a magnetic flux tube triggered by an orbiting planet. 
The simulated light curve changes as a function of the planet position inside the stellar magnetosphere.
Different sizes of the exoplanet orbital radius have been analysed, 
the corresponding values of planetary revolution periods are reported in the top x-axis of the figures. 
The simulations have been performed assuming two different values of the thickness of the flux tube,
$\Delta=90^{\circ}$ (top panels) and $\Delta=30^{\circ}$ (bottom panels). 
The dashed and the dot-dashed horizontal lines represent, respectively, the left- and right-handed circularly polarised
pulse components detected at 8.44 GHz.
%{\bf I 6 PANNELLI IN BASSO SONO STATI OTTENUTI ASSUMENDO UN PIANO DI RIVOLUZIONE INCLINATO DI 30 GRADI RISPETTO
%AL PIANO EQUATORIALE STELLARE}
} 
\label{season} 
\end{figure*}
%============================================fig 6

%============================================fig 6
\begin{figure*}
%\resizebox{\hsize}{!}{\includegraphics{fig6.eps}}
\resizebox{\hsize}{!}{\includegraphics{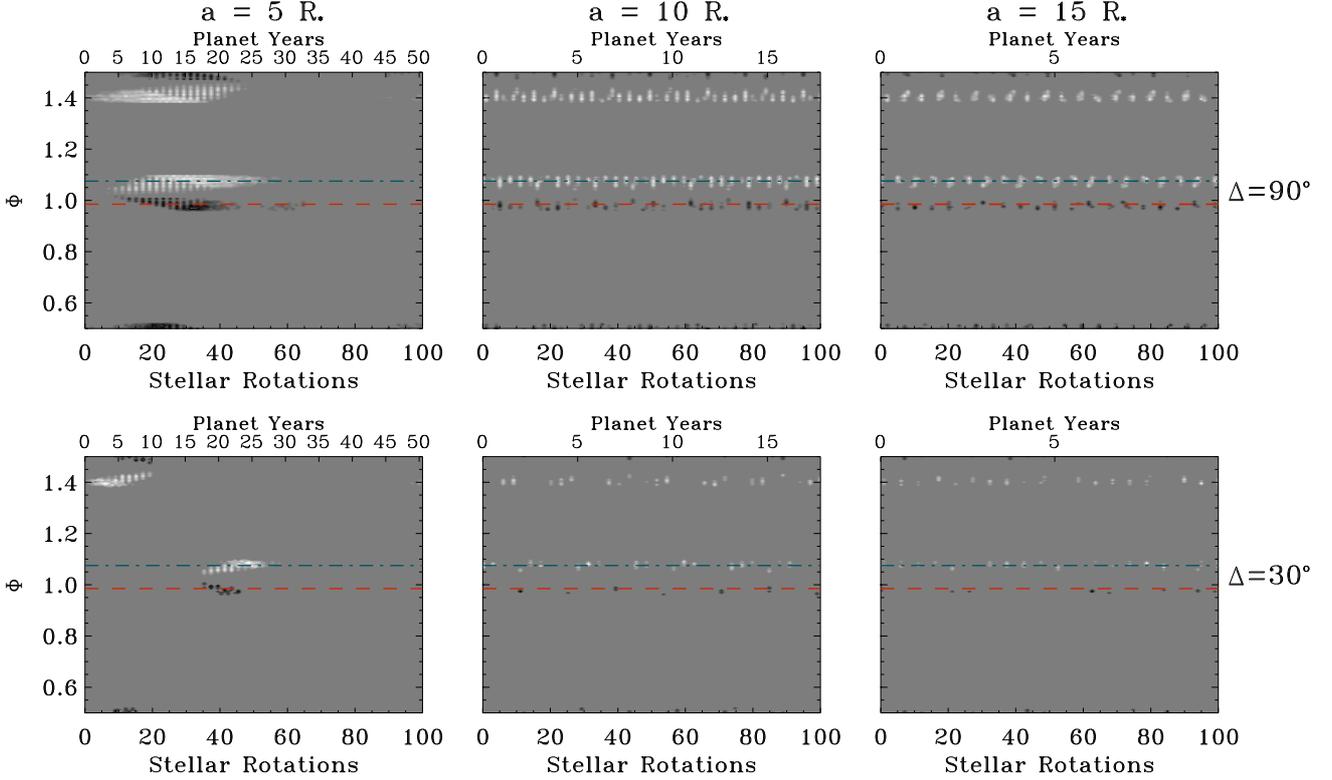}}
\caption{
{Caption similar to the case of Fig.~\ref{season}. The simulations pictured in figure were obtained assuming an orbiting plane
inclined of $30^{\circ}$ respect to the star rotation plane.}
} 
\label{season_inclined} 
\end{figure*}
%============================================fig 6

%============================================fig 7
\begin{figure*}
\resizebox{\hsize}{!}{\includegraphics{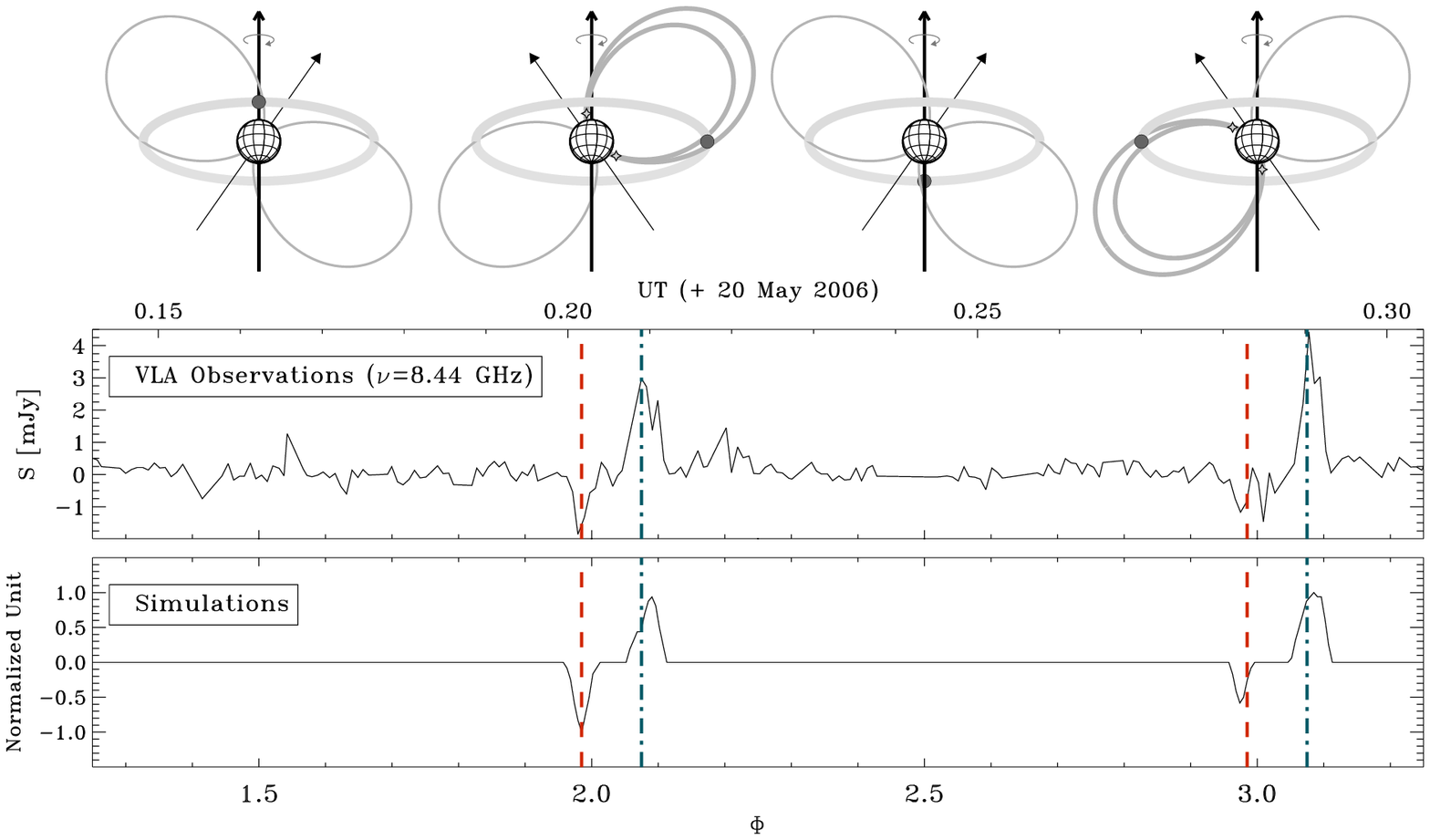}}
\caption{Top panels: cartoon showing the magnetospheric orientation of TVLM\,513-46546
and the corresponding planet orbital position able to trigger the stellar auroral radio emission 
beamed along directions observable from the Earth.
Middle panel: light curve of TVLM\,513-46546 observed at 8.44 GHz lasting 2 consecutive stellar rotations.
Bottom panel: two consecutive simulated auroral light curves captured from an active season  
of the star. The simulations of these light curves have been carried out assuming the
closer orbital configuration and the slimmer flux tube size; simulations shown on the bottom left panel of Fig.~\ref{season}.
The dashed and the dot-dashed vertical lines individuate, respectively, the left- and the right-handed circularly polarised
pulse components detected at 8.44 GHz.
{The cartoon and the simulations 
refer to the case of planet orbital plane coinciding with the stellar rotation plane.}
}
\label{cartoon}
\end{figure*}
%============================================fig 7

{

%Could be able to explain 
%To reproduce the timing and the polarised features of the X band auroral light curve,
%In the following
We analyse if the auroral radio emission induced by the star-planet interaction
%a scenario like the Io-Jupiter interaction %, where the Io-DAM emission originates in the Io's flux tube. In fact,}
could be applied to the case of TVLM\,513-46546.
\citet{hess_zarka11} have extensively analysed %the observable signatures of 
the auroral radio emission induced by the star planet interaction 
%of a star with one of its orbiting exo-planets 
as a tool to derive information regarding the geometry of the system.
The authors take into account two possible origins of the auroral radio emission: % were examined \citep{hess_zarka11};
the case of auroral radio emission from the star induced by the exoplanet 
(analogous to the case of  auroral radio emission in Jupiter, controlled by the moon Io) and
the case of emission from the magnetosphere of the exoplanet induced by the stellar wind.
The high-frequency cutoff of the auroral radio emission is directly related to the magnetic-field strength
at the surface of the object where it takes place, be it the star or the planet.
The magnetic-field strength of exoplanets  is expected to be comparable to the magnitude of the magnetic-field strength of Jupiter (few Gauss).
%lower then few Gauss, 
%the study of 
The auroral radio emission from the exoplanet, powered by the wind of the parent star \citep{nichols_milan16},
%is a very interesting since topic  \citep{nichols_milan16} that 
can be explored in the low-frequency radio domain.
%can be explored at the MHz frequency window.
Because of the strength of the magnetic-field 
deduced from the analysis of  the auroral radio emission in TVLM\,513-46546 (at the kGauss level),
%highest frequency ($\nu=8.44$ GHz) of the 
the case of the auroral radio emission arising from the stellar flux tube crossing the exoplanet seems more plausible.}
%if it is assumed that there is an auroral component originating from the star planet interaction.}
This scenario is shown in Fig.~\ref{sezione}, where it is possible to distinguish two auroral cavities.
The poleward cavity
{where the 4.88 GHz originates is related to the centrifugally-opened field lines.}
%filled by ionised material, that is not rigidly corotating with the star.}
%The poleward cavity is associated with the open field lines. 
Far from the star, these lines locate the system of currents generated by the centrifugal force \citep{nichols_etal12}.
The equatorward cavity, {responsible of the 8.44 GHz auroral emission}, 
is instead related to the flux tube excited by the planet.
The combined auroral emissions arising from the flux tubes crossing the planet and from the auroral cavity
involving the whole magnetosphere {could be}
%will be 
responsible for the different auroral features observed {at 4.88 and 8.44 GHz} in TVLM\,513-46546.

{
%As already pointed out, 
The stellar auroral radio emission triggered by an 
external body is modulated by the orbital period of the planet \citep{hess_zarka11}. 
To check if the planet-induced auroral radio emission could be a plausible scenario for TVLM\,513-46546, we}
performed simulations of the auroral radio emission arising only from the flux tube excited by
a small body orbiting close to the star, not from the full auroral ring.
{The frequency of the simulated auroral radio emission is equal to 8.44 GHz, 
and the field topology of the small scale stellar magnetic loops was approximated by a simple dipole.}
%and, even if this is an approximation, the field topology of the small scale stellar magnetic loops was assumed as a simple dipole.}
%and the magnetic field topology was assumed as a simple dipole.}
%This scenario is shown in Fig.~\ref{sezione}, where it is possible to distinguish two auroral cavities.
%The poleward cavity is associated with the open field lines. Far from the star,
%these lines locate the system of currents generated by centrifugal force \citep{nichols_etal12}.
%The equatorward cavity is instead related to the flux tube excited by the planet.
%The combined auroral emissions arising from the flux tubes crossing the planet and from the auroral cavity
%involving the whole magnetosphere {\bf could be}
%%will be 
%responsible for the different auroral features observed in TVLM\,513-46546. 
The upper limit to the orbital radius ($a$) of the planet  is assumed less or equal to the minimum  
magnetospheric size adopted in the previous simulations (15 R$_{\ast}$).
This is to take into account the results obtained by \citet{kuznetsov_etal12} who conclude that an external body 
$\approx 24$ stellar radii away from TVLM\,513-46546 would not be able
to explain the timing of its auroral radio emission.
The analysed values of the planet orbital radius $a$ are: 5, 10, and 15 stellar radii.
{Two possible inclinations of the orbital plane have been taken into account:
the case of a planet orbiting in the plane defined by the stellar rotation and 
the case of a planet with orbital plane inclined of $30^{\circ}$.}

The stellar mass of TVLM\,513-46546 is about 0.07 M$_{\odot}$ \citep{dahn_etal02}.
Applying the third Kepler's law, the chosen values of the orbital radii give 
planet orbital periods of 1.98, 5.61, and 10.31 times the stellar rotation period, respectively ($P_{\mathrm {rot}} \approx 1.96$ hours).
For the magnetic sector of the active flux tube,
two cases have been analysed. In the first one, the angle at the centre ($\Delta$) of the magnetic sector
is fixed equal to $90^{\circ}$, similar to the longitudinal extension of the tail following the spot of Io's induced aurora, 
as observed in the ultraviolet images of Jupiter \citep{clarke_etal02}.  In the other case, the angle $\Delta$ 
is assumed equal to $30^{\circ}$.
We adopt $\beta=120^{\circ}$ (central value of the estimated range given in Table~\ref{sol_simul}, 
equivalent to a dipole absolute tilt of $60^{\circ}$).
The corresponding parameters that control the auroral beam pattern ($\delta$ and $\theta$) are reported in Table~\ref{sol_simul}.
We discard the hypothesis $\beta=360^{\circ}-120^{\circ}=240^{\circ}$, because this assumption implies a too large $\Delta \Phi$ value.

For the cases analysed here, we 
simulated the light curves as a function of the planet orbital position. 
Such simulated dynamic auroral light curves are shown in {Figs.~\ref{season} and \ref{season_inclined}.}
In the case of auroral emission induced in the thicker flux tube ($\Delta=90^{\circ}$),
we find a high occurrence of the phenomenon; {see the top panels of Fig.~\ref{season} and Fig.~\ref{season_inclined}.}
In particular, {in the case of the planet orbital plane coinciding with the stellar rotation plane,} 
during 100 consecutive stellar rotations the pulses are detectable in many stellar periods.
The percentage of periods characterised by at least one pulse are: 
82\% assuming planet orbital radius $a=5$ R$_{\ast}$ (82\% RCP, 74\% LCP); 
91\% with $a=10$ R$_{\ast}$ (91\% RCP, 89\% LCP);
80\% with $a=15$ R$_{\ast}$ (78\% RCP, 77\% LCP).
{The auroral pulse detection rate, estimated for the case of 
the orbital plane inclined of $30^{\circ}$, are instead:
78\% with $a=5$ R$_{\ast}$ (68\% RCP, 60\% LCP); 
91\% with $a=10$ R$_{\ast}$ (91\% RCP, 69\% LCP);
80\% with $a=15$ R$_{\ast}$ (78\% RCP, 62\% LCP).}
In the case of auroral radio emission arising from a planet flux tube followed by a long tail, the simulations generally
predict light curves characterised by the occurrence of two pulses (doubly or singly peaked) per stellar rotation.
An exception to this are  the simulations in which the planet is closer to the star (5 R$_{\ast}$),  predicting periods during
which we observe two pulses per stellar rotation, periods with one pulse per rotation, and quiet periods.
Such seasonal periodicity of the pulse occurrence is clearly highlighted only in this case.
The analysis of the occurrence of the auroral emission simulated in the case of the slimmer flux tube ($\Delta=30^{\circ}$),
{bottom panels of Fig.~\ref{season} and Fig.~\ref{season_inclined}}, shows a lower detection rate.
{In the case of orbital plane coinciding with the stellar rotation plane the detection rates are:}
43\% with $a=5$ R$_{\ast}$ (41\% RCP, 35\% LCP); 
49\% with $a=10$ R$_{\ast}$ (36\% RCP, 32\% LCP);
42\% with $a=15$ R$_{\ast}$ (31\% RCP, 27\% LCP).
{Whereas in the case of orbital plane inclined of $30^{\circ}$, the detection rates are:
41\% with $a=5$ R$_{\ast}$ (41\% RCP, 18\% LCP); 
47\% with $a=10$ R$_{\ast}$ (37\% RCP, 16\% LCP);
36\% with $a=15$ R$_{\ast}$ (25\% RCP, 16\% LCP).}
This happens because in the case of the thick flux tube ($\Delta=90^{\circ}$)
it is possible to have a number of planet orbital positions able to cause a detectable auroral radio emission larger than that
obtainable in the case of a thin flux tube ($\Delta=30^{\circ}$). 
We also note that, for the adopted stellar geometry, the detection of right-hand circularly polarised (RCP) pulses is favoured
for every combination of the model parameters chosen to characterise the planet-induced auroral emission.
{These conclusions hold for both the considered orbital plane inclinations,
even though, by assuming the inclined orbital plane, we simulate a lower occurrence of the auroral radio emission.
%In fact, the misalignment between the plane of the magnetic equator and the orbital plane causes the changes of the planet magnetic latitude, 
%while the star and the planet rotate. The $L$-shell parameter of the planet flux tube is then a function of the planet orbital position.
%For the simulated stellar geometry ($|\beta|=60^{\circ}$), the further inclination of $30^{\circ}$ of the orbital plane
%has the consequence that there exist star-planet reciprocal positions favorable for the detection of the auroral pulses, 
%but that are characterised by the magnetic axis lying in the planet orbital plane, 
%follow that the magnetic field line crossed by the planet coincides with the magnetic axis.
%The magnetic field lines coinciding with the polar axis are closed at distance infinite, 
%and the corresponding size of the auroral cavity vanishes.
%\textcolor{red}{chiarire meglio questo concept}{\it follow that $L=\infty$, and the corresponding auroral cavity vanishes.}
}

In the case of the slimmer flux tube, the simulated auroral radio emission looks like a periodic phenomenon.
The simulations predict two pulses per spin period, occurring
every $\approx 3$ stellar rotations, in the case of a planet orbiting 10 R$_{\ast}$ far from the star,
and every $\approx 5$ stellar rotations, in the case of a planet orbiting 15 R$_{\ast}$ far from the star.
In these two cases, the planet covers  a full orbit every 5.61 and 10.31 stellar rotation periods, respectively,
and it can take the appropriate position twice per stellar period. 
When the planet is closest to the star (orbital radius equal to 5 R$_{\ast}$ and
revolution period about 3 hours and 52 minutes),
we simulate a clear seasonal periodicity of the pulse occurrence {(bottom left panels of Fig.~\ref{season} and Fig.~\ref{season_inclined})},
similarly to the thicker flux tube case.
Even in this case the seasonal modulation of the planet-induced auroral radio emission predicts active  and
quiet periods, without observable pulses.
In the case of a close planet, the active periods are characterised by light curves with
a single pulse. The adjacent active phases show peaks of reversed polarisation signs
(right-left instead of left-right and vice versa). 
The active phase taken into exam shows a left-right pulse (Fig.~\ref{tvlm_dic_spec} top panel), while
in another epoch a right-left pulse has also been found in TVLM\,513-46546
(measurements in C band performed with the Arecibo radiotelescope; \citealp{kuznetsov_etal12}).
On the basis of these observed features,
the planet-induced TVLM\,513-46546 auroral emission is well reproduced, among the chosen orbital radii,
by the closest planet configuration: $a=5$ R$_{\ast}$. 
To quantify the flux tube angular size, a systematic search of the 
auroral radio emission detection rate is needed.

{Similarly to the Sun-like magnetic cycle, 
it has been theoretically predicted that also in the case of
 fully-convective slowly-rotating late-type stars 
the magnetic-field axis periodically changes its orientation, 
reverting the north to south magnetic polarity \citep{yadav_etal16}.
%The circularly polarisation sign of the stellar auroral radio emission is directly related
%to the magnetic polarity of the regions where it took place.
%The right hand circularly polarised (RCP) pulses are related to the north magnetic hemisphere,
%conversely, the left hand polarisation state (LCP) is associated to the auroral pulses arising from the southern stellar hemisphere.
%%The possible existence of solar like magnetic activity cycles also in that case of the fast rotating UCDs 
%%was assumed as explanation of the detections in different epochs of  
Such possible magnetic-polarity reversal, acting also in the case of the fast rotating UCDs, was claimed
as the origin of the detection in different observing epochs of auroral pulses with opposite polarisation sense \citep{route16}.
Our analysis highlights that 
the planet-induced stellar auroral radio emission scenario is also
able to explain 
this observational evidence.} % is also explained within the scenario of the planet induced auroral radio emission. }

The planet-induced simulated auroral light curves
covering two consecutive rotation periods are displayed  in Fig.~\ref{cartoon} (bottom panel),
{the simulations shown in  the figure refer to the case of the planet orbital plane coinciding with the stellar rotation plane.}
These can be compared with the observed 8.44 GHz light curve (middle panel)
and with the corresponding stellar magnetosphere orientation and planet orbital position (top panel). 
Despite the favourable orientation of the stellar magnetosphere,
the planet-induced auroral pulses are not detected
close to the rotational phases $\Phi \approx 1.5$ and $\Phi \approx 2.5$;
see Fig.~\ref{cartoon} top panels.
This is because the flux tube crossed by the planet does not lie on the plane of the sky.
At phases $\Phi \approx n \times 1$ ($n$ integer), this flux tube is instead roughly placed on the sky plane and 
the planet-induced auroral radio emission is then observable. 
The planet orbital revolution period is about twice the stellar rotation period. Therefore,
at every stellar rotation the planet 
covers about half of its orbit. This implies that the previous planet flux tube orientation repeats at 
every full stellar rotation.

{
To search for a companion orbiting around TVLM\,513-46546, observations were 
conducted using the Very Long Baseline Interferometry (VLBI) technique.
In particular, measurements have been performed with
the Very Long Baseline Array (VLBA) \citep*{forbrich_berger09,forbrich_etal13}
and with the European VLBI Network (EVN) \citep{gawronski_etal16} interferometers.
The VLBI measurements  rule out the case of the existence of a companion as massive as Jupiter or more,
with a long orbital period (years). 
Unfortunately, the VLBI astrometric sensitivity was not enough to confirm or rule out the possible existence of
a low mass exoplanet orbiting close to the star.

}

%________________________
\section{Summary and outlook}
\label{summ}

In this paper, we have presented the simulation of the fully-polarised radio pulses
of the well-known UCD TVLM\,513-46546, observed by \citet{hallinan_etal07} in May 2006.
We have used the 3D model introduced in \citet{leto_etal16} to reproduce the pulse profile
of the auroral radio emission arising from a dipolar cavity.
{In this paper, this model has been improved
%we describe in this paper a modified model 
to simulate the auroral radio
emission from a magnetosphere shaped like an offset-dipole.
Our analysis 
has been able to investigate the 
magnetic-field topology of the magnetosphere in TVLM\,513-46546, giving an indirect estimation of its strength. We have also been able to}
%\textcolor{red}{non si capisce che vuole dire}{\it was able to probe} the complex stellar magnetic field topology, and  to}
%The analysis performed in this paper gives indirect
%hints about the stellar geometry and the magnetic field topology,
%that could be confirmed by the direct measurement of the  TVLM\,513-46546 effective magnetic field curve.
%Moreover, we 
infer some constraints on the number density
of the thermal electrons trapped {inside}. %within the stellar magnetosphere.

{The scenario of the stellar auroral radio emission induced by an orbiting planet has also been examined.}
%{\it RENDI MENO FERMA QUESTA AFFERMAZIONE. The most significant result achieved by this study is the
%hint to the existence of a close-in orbiting planet.}  
By interacting
with the magnetic field of TVLM\,513-46546, this planet would induce ECM emission
within the  crossed flux tube, like the interaction of  Jupiter's magnetosphere with its moons.
{The simulations of the auroral light-curve time evolution is compatible with the observed one,
suggesting that the planet-induced auroral radio emission is a plausible scenario  to produce  
some spectral components of the auroral radio emission in TVLM\,513-46546.}
If the existence of such a planet  is confirmed, 
the study of the temporal evolution of the stellar auroral radio emission would be a powerful tool 
for the detection of exoplanets.
{As already pointed out by \citet{hess_zarka11}}, 
it is necessary to perform long-term monitoring of the stellar auroral radio emission to acquire more information about the  orbital parameters of the planet.

Although in the last few years the improvement of observational capabilities in the infrared
allowed the detection of a number of UCDs,
the number of these stars in which radio emission has been detected is still limited 
by their intrinsic weakness as radio sources.
It is thus likely that the sensitivity of the existing instruments in many cases does not allow us to detect radio emission from UCDs, 
which introduces a strong selection bias in our knowledge of the radio emission of  very late-type main sequence stars.

With its unprecedented sensitivity and field of view, the Square Kilometre Array (SKA) will be an ideal instrument for deep surveys.
Searching for radio emission from UCDs, SKA will extend the sample population 
and allow the detection of low flux density rotational modulation (few percent  in amplitude), which is
crucial to assess the stellar geometry.  
Also,  the ability of the SKA to make simultaneous multi-frequency measurements with great sensitivity
will make it possible to study the temporal and spectral evolution of the auroral radio emission observed
in UCDs. In fact, only using simultaneous multi-frequency measurements it should be possible
to disentangle the contributions to the auroral radio emission  from the open field lines and from the possible planet flux tube.

In addition, the SKA will open a window onto radio frequencies previously 
unexplored at high sensitivity. This makes SKA the most powerful instrument to hunt for stellar and planet
coherent emission in the radio sky. Following the ECM theory,  the amplified frequency escaping from
the magnetosphere is directly related to the magnetic field strength \citep{wu_lee79,melrose_dulk82}. For example, pulses attributed to
ECM emission have been observed at kHz and MHz frequencies from the magnetised planets of our solar system \citep{zarka98}.  
We put forward that this will open up the catalogue of potential candidates for detection of such kind of emission not only 
through the population of UCDs, but also among the exoplanets orbiting around
stars of spectral type similar to our Sun.
The possible exoplanetary auroral radio emission can be triggered by the interaction of the exoplanet magnetosphere with the 
radiatively-driven wind arising from the parent star \citep{zarka07}.
If we search for auroral radio emission from exoplanets with polar strengths similar to those of the magnetised planets in the solar system,
the ECM emission process will operate
in a frequency range that only SKA will cover with high sensitivity and resolution.

\appendix

\section{Offset dipole: procedures}
%\section{Procedures}
\label{offset_dip}

{
The orientation of a magnetosphere shaped like an oblique rotator is defined by
the rotation axis inclination respect to the line of sight (angle $i$), 
the tilt angle of the magnetic axis respect to the rotation axis ($\beta$), and
the rotational phase ($\Phi$).
In the case of a central dipole,
the visibility of the auroral radio emission 
as a function of the stellar geometry was analysed in Paper I. 
In the following, we summarise the procedures used in this paper to simulate the auroral radio emission 
arising from a stellar magnetosphere shaped like a non-central dipole.

%First of all we define the reference frames where the 

%The magnetic field vector components are calculated in the $Oxyz$ reference frame, anchored to the star,
%having the $z$-axis parallel to the dipole axis, and the $x$-axis located in the plane 
%passing through the $z$-axis  and containing the rotation axis. 

To calculate the magnetic field vector generated by an offset-dipole,
the magnetic dipole moment $\vec{m}$ %(magnitude $m=\frac{1}{2}B_{\mathrm p} R_{\ast}$) 
is located in a generic grid point ($O^{\prime}$) within the star, identified by its spatial coordinates $(\delta x, \delta y, \delta z)$
in the native reference frame $Oxyz$, where the origin $O$ coincides with the centre of the star,
the $z$-axis is parallel or coincident with the dipole axis, and the $x$-axis lies into the plane parallel to the $z$-axis
and containing the rotation axis.

In the reference frame anchored with the offset-dipole,
the new coordinates $(X,Y,Z)$ of the grid points 
%in the reference frame anchored with the offset-dipole,
are defined by the simple coordinates translation $(x+\delta x, y+\delta y, z+\delta z)$.
In the translated $O^{\prime}XYZ$ reference frame, the $Z$-axis coincides with the dipole axis and the $X$-axis is located in the plane 
passing through the $Z$-axis and parallel to (or containing) the rotation axis. 
The corresponding magnitude of the magnetic dipole moment is defined as follows:
\begin{displaymath}
m=\frac{1}{2}B_{\mathrm p} (R_{\ast}-\delta r_{\mathrm Z})
\end{displaymath}
where $\delta r_{\mathrm Z}$ is the distance of the offset dipole from the stellar surface 
along the direction parallel to $Z$-axis,
and $B_{\mathrm p}$ is the polar magnetic field strength, that was fixed equal to 3000 Gauss.

The three components $(B_{\mathrm X},B_{\mathrm Y},B_{\mathrm Z})$ of the magnetic field vector are
calculated in each grid point taking into account the non-central position
of the magnetic dipole moment. 
In the new reference frame $O^{\prime} X Y Z$,
the magnetic vector components are easily calculated 
using the equations of the simple dipole
(see Appendix A.2 of \citealp{trigilio_etal04}).

%is convenient to calculate the
%magnetic vector components using
%the equations of the simple dipole, and
%the dipolar magnetic field lines, defined by using the equation $r^{\prime}=L \cos ^2 \lambda$,
%where $r^{\prime}$ is the distance of the grid points to the origin $O^{\prime}$ (where $\vec{m}$ is located).
%In this new reference frame,
%anchored with the grid point where is located the magnetic dipole moment,
%and having the axis parallel to the $x$, $y$ and $z$ axes,
%the dipolar magnetic field lines were defined by using the equation $r^{\prime}=L \cos ^2 \lambda$,
%where $r^{\prime}$ is the distance of the grid points to the $\vec{m}$ location.
%%and $\lambda$ the magnetic latitude in the offset reference frame.
The magnetic vectorial field %, calculated in the reference frame anchored with the star 
is then rotated in the observer reference frame $Ox^{\prime}y^{\prime}z^{\prime}$,  
where the origin $O$ coincides with the centre of the star, the $x^{\prime}$-axis is parallel to the line of sight,
and the plane $y^{\prime}z^{\prime}$ coincides with the plane of the sky.
The method was described in the Appendix A.3 of \citet{trigilio_etal04}.

In the reference frame ($O^{\prime}XYZ$) anchored with the offset-dipole,
it is also possible to locate the grid points falling inside the auroral cavity using the equation
of the dipolar magnetic field lines, defined by using the equation $r^{\prime}=L \cos ^2 \lambda$,
where $r^{\prime}$ is the distance of the grid points to the origin $O^{\prime}$ (where $\vec{m}$ is located).
Following the method described in Paper I, it is then possible to find 
those grid points that radiate auroral radio emission  
within a beam pattern that intercepts the line of sight
and that are tuned at the assigned simulation radio frequency $\nu$.
To take into account the non-central position of the magnetic-dipole moment ($O^{\prime} \neq O$),
the distance to the centre of the star from the grid points 
that are able to radiate detectable auroral radio emission
%the grid points that are able to radiate detectable auroral radio emission
%have to satisfy the condition that their distance from the centre of the star
has to be larger than the stellar radius.
The star shadowing effect has also been taken into account.  
}

\section{Offset dipole: simulations}
%\section{Simulations}
\label{app}

\begin{figure*}
\resizebox{\hsize}{!}{\includegraphics{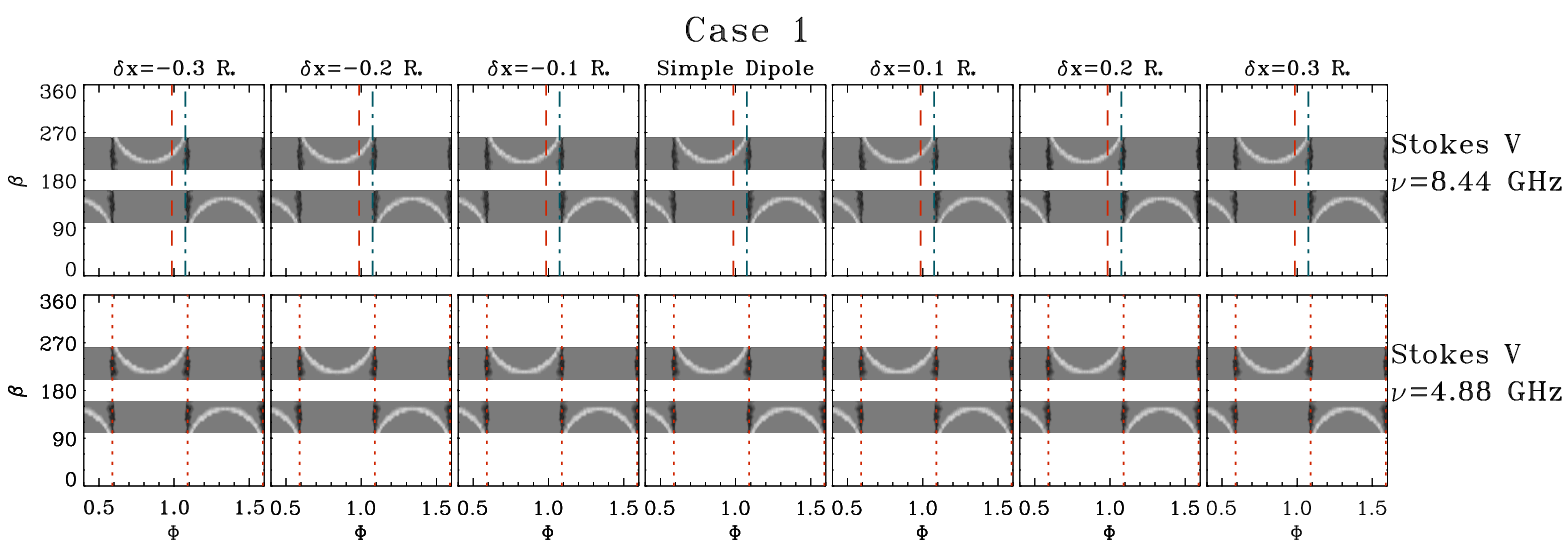}}
\resizebox{\hsize}{!}{\includegraphics{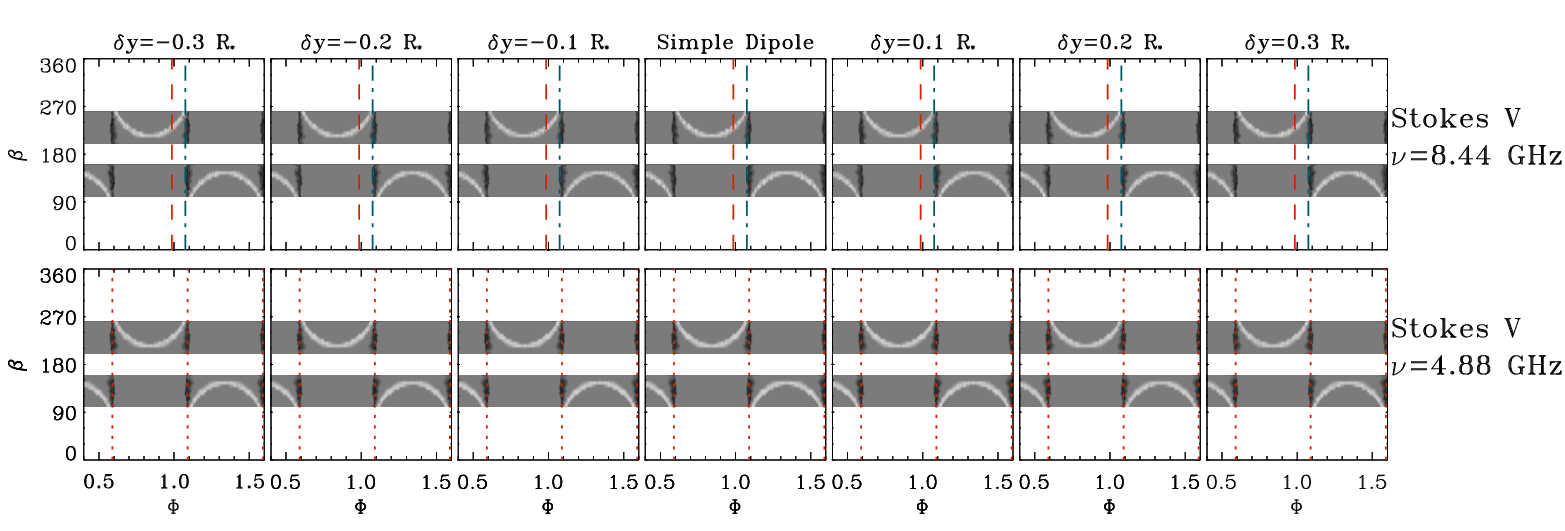}}
\resizebox{\hsize}{!}{\includegraphics{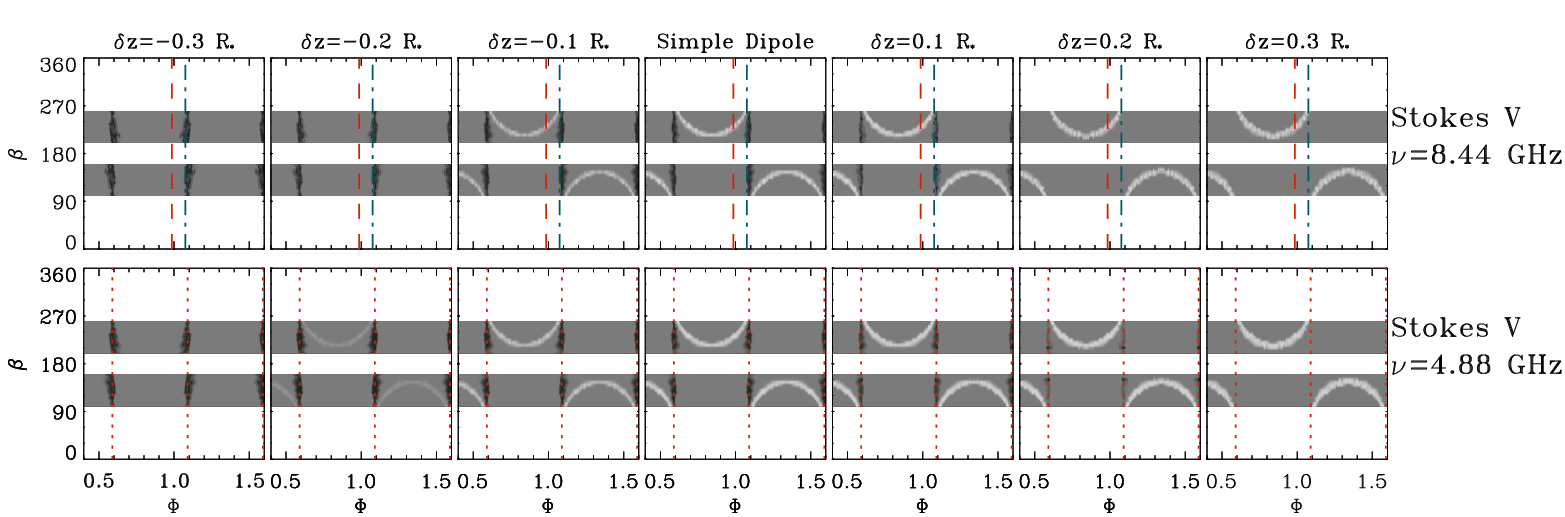}}
\caption{Simulations of the dynamical auroral light
curves at the X and C bands (8.44 and 4.88 GHz respectively),
similar to the simulations from a simple dipole shown in Fig.~\ref{tvlm_simulaz}. 
The adopted sets of the model parameters are listed in Table~\ref{sol_simul}.
The auroral radio emission simulations here displayed explore the effects of
the dipole shift along the $x$, $y$ and $z$-axes.
The simulations shown in figure refer to the Case 1 described in Sec.~\ref{cso}.
}
\label{case1_4pole} 
\end{figure*}

\begin{figure*}
\resizebox{\hsize}{!}{\includegraphics{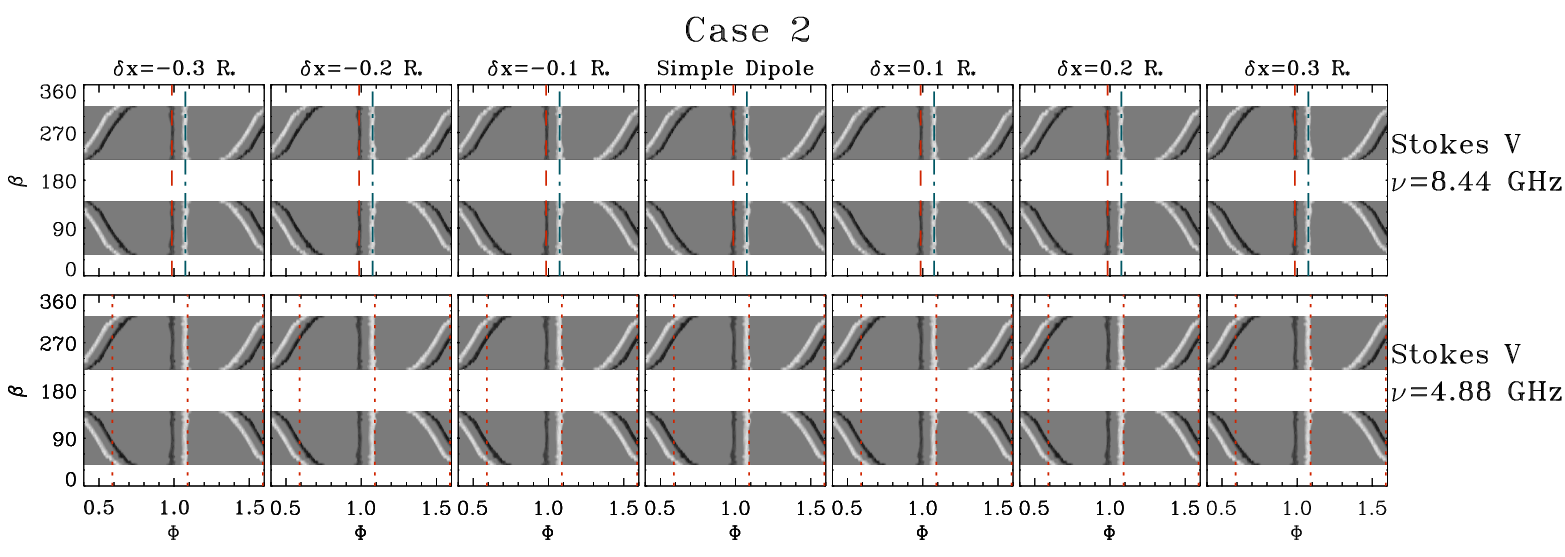}}
\resizebox{\hsize}{!}{\includegraphics{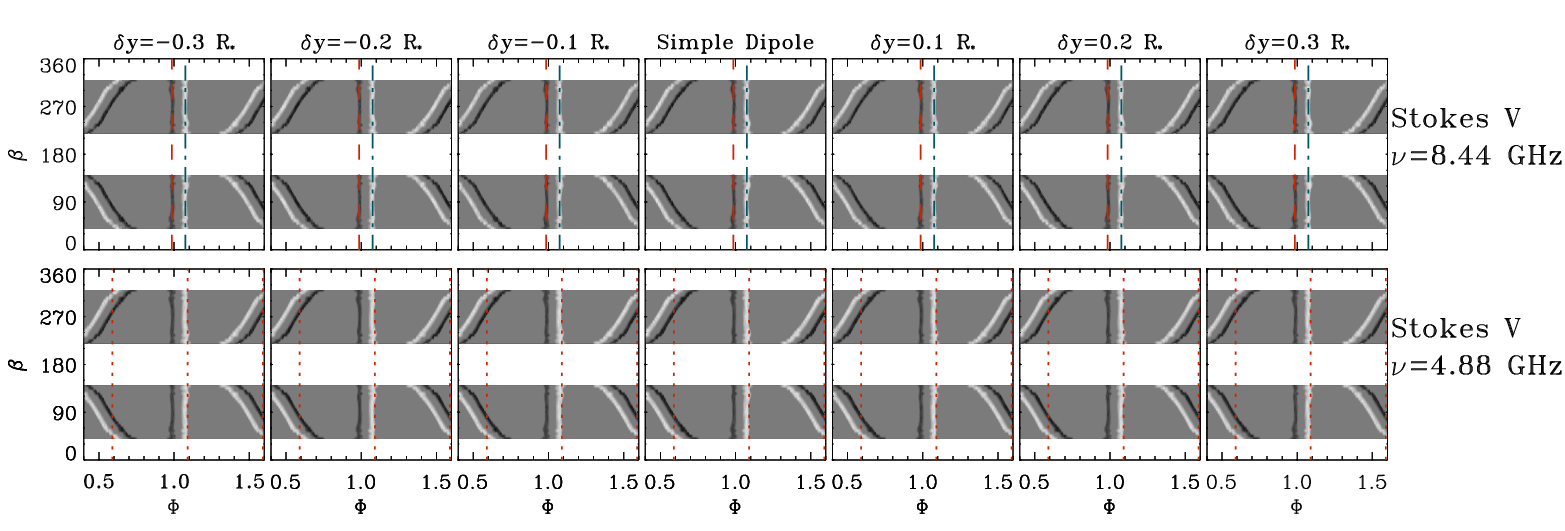}}
\resizebox{\hsize}{!}{\includegraphics{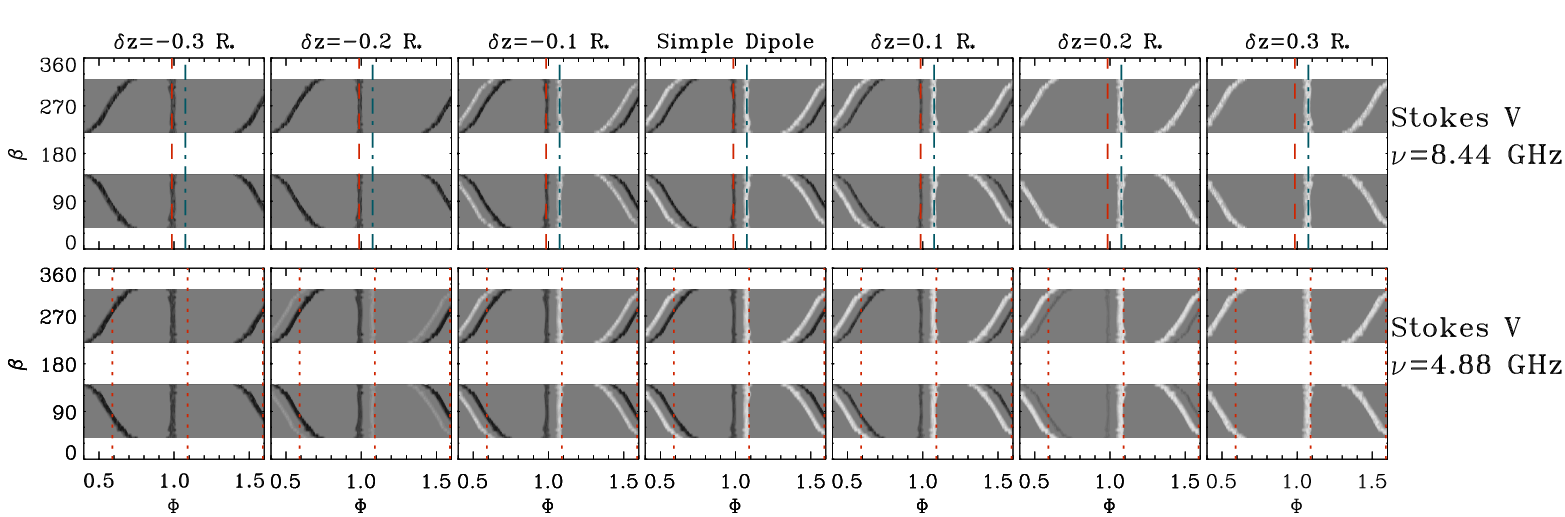}}
\caption{Same caption of Fig.~\ref{case1_4pole},
but simulations referred to the Case 2 described in Sec.~\ref{cso}.
}
\label{case2_4pole} 
\end{figure*}

{
The simulations of the auroral radio emission visibility 
from a non central dipolar field were performed assuming the
model parameters listed in Table~\ref{sol_simul}.
The effects of the dipole shift along the $x$, $y$ and $z$-axes were analysed.
The dipole was shifted along the $xyz$-axes
in the range from $-0.3$ to 0.3 [R$_{\ast}$], with a step of 0.1 [R$_{\ast}$].
The simulated light curves, referring to the two cases analysed in Sec.~\ref{cso}, 
are shown in Figs.~\ref{case1_4pole} and \ref{case2_4pole}
as a function of the dipole shift value.

}

%============================================fig 6
%============================================fig 6

%============================================fig 6
%============================================fig 6

\section*{Acknowledgments}
We thank Dr. Sebastien Hess for his useful criticisms that helped to improve the paper.
This paper includes archived data obtained through the Karl G. Jansky Very Large Array Online Data Archive
(\url{https://archive.nrao.edu/archive/advquery.jsp}), operated by the 
National Radio Astronomy Observatory\footnote{The National Radio Astronomy Observatory is a facility of the National Science Foundation operated under cooperative agreement by Associated Universities, Inc.} (NRAO).
This research has made use of the SIMBAD database, operated at CDS, Strasbourg, France.

%________________________________________________________________


\begin{thebibliography}{99}
\bibitem[\protect\citeauthoryear{Antonova et al.}{2008}]{antonova_etal08} 
	Antonova A., Doyle J.G., Hallinan G., Bourke S., Golden A., 2008, A\&A, 487, 317
\bibitem[\protect\citeauthoryear{Antonova et al.}{2013}]{antonova_etal13} 
	Antonova A., Hallinan G., Doyle J.G., Yu S., Kuznetsov A., Metodieva Y., Golden A., Cruz K.L., 2013, A\&A, 549, 131
\bibitem[\protect\citeauthoryear{Babcock}{1949}]{babcock49} 
	Babcock H.W., 1949, Observ. 69, 191
\bibitem[\protect\citeauthoryear{Benz \& Guedel}{1994}]{benz_guedel94} 
	Benz A.O., Guedel M.,  1994, A\&A, 285, 621
\bibitem[\protect\citeauthoryear{Berger et al.}{2001}]{berger_etal01} 
	Berger E., Ball S., Becker K.M., et al., 2001, Nature, 410, 338	
\bibitem[\protect\citeauthoryear{Berger}{2002}]{berger02} 
	Berger E., 2002, ApJ, 572, 503
\bibitem[\protect\citeauthoryear{Berger}{2006}]{berger06} 
	Berger E., 2006, ApJ, 648, 629	
\bibitem[\protect\citeauthoryear{Berger et al.}{2008a}]{berger_etal08a} 
	Berger E., Gizis J.E., Giampapa M.S., et al., 2008a, ApJ, 673, 1080 
	%Rutledge R.E., Liebert J., Mart\'{i}n E., Basri G., Fleming T.A., et al., 2008a, ApJ, 673, 1080
\bibitem[\protect\citeauthoryear{Berger et al.}{2008b}]{berger_etal08b} 
	Berger E., Basri G., Gizis J.E., et al., 2008b, ApJ, 676, 1307	
\bibitem[\protect\citeauthoryear{Berger et al.}{2010}]{berger_etal10} 
	Berger E., Basri G., Fleming T.A., et al., 2010, ApJ, 709, 332 
	%Giampapa M.S., Gizis J.E.,  Liebert J., Mart\'{i}n E., Phan-Bao N., et al., 2010, ApJ, 709, 332	
\bibitem[\protect\citeauthoryear{Bonfond et al.}{2012}]{bonfond_etal12} 
	Bonfond B., D. Grodent D., Gerard J.-C., et al., 2012, GeoRL, 39, 1105	
\bibitem[\protect\citeauthoryear{Burgasser \& Putman}{2005}]{burgasser_putman05} 
	Burgasser A.J., Putman M.E., 2005, ApJ, 626, 486	
\bibitem[\protect\citeauthoryear{Burgasser et al.}{2015}]{burgasser_etal15} 
	Burgasser A.J., Melis C., Todd J., Gelino C.R., Hallinan G., Gagliuffi D.B., 2015, AJ, 150, 180	


\bibitem[\protect\citeauthoryear{Cecconi et al.}{2012}]{cecconi_etal12} 	
	Cecconi B., Hess S., H\'{e}rique A., et al., 2012, P\&SS, 61, 32

\bibitem[\protect\citeauthoryear{Chandra et al.}{2015}]{chandra_etal15} 	
	Chandra P., Wade G.A., Sundqvist J.O., et al., 2015, MNRAS, 452, 1245
\bibitem[\protect\citeauthoryear{Chabrier \& Kuker}{2006}]{chabrier_kuker06} 
	Chabrier G, Kuker M., 2006, A\&A, 446, 1027
\bibitem[\protect\citeauthoryear{Clarke et al.}{2002}]{clarke_etal02} 	
	Clarke J.T., Ajello J., Ballester G., et al., 2002, Nature, 415, 997
	
\bibitem[\protect\citeauthoryear{Christensen, Holzwarth \& Reiners}{Christensen et al.}{2009}]{christensen_etal09} 
	Christensen U.R., Holzwarth V., Reiners A., 2009, Nature, 457, 167
	
\bibitem[\protect\citeauthoryear{Dahn et al.}{2002}]{dahn_etal02} 	
	Dahn C.C., Harris H.C., Vrba F.J., et al., 2002, AJ, 124, 1170
\bibitem[\protect\citeauthoryear{Donati et al.}{2006}]{donati_etal06} 	
	Donati J.-F., Forveille T., Collier Cameron A., et al., 2006, Sci, 311, 633
\bibitem[\protect\citeauthoryear{Donati et al.}{2008}]{donati_etal08} 
	Donati J.-F., Morin J., Petit P., et al., 2008, MNRAS, 390, 545
\bibitem[\protect\citeauthoryear{Doyle et al.}{2010}]{doyle_etal10} 
	Doyle J.G., Antonova A., Marsh M.S., Hallinan G., Yu S., Golden A., 2010, A\&A, 524, 15
\bibitem[\protect\citeauthoryear{Drake et al.}{1987}]{drake_etal87} 
	Drake S.A., Abbot D.C., Bastian T.S., Bieging J.H., Churchwell E., Dulk G., Linsky J.L, 1987, ApJ, 322, 902
\bibitem[\protect\citeauthoryear{Ergun et al.}{2000}]{ergun_etal00}
	Ergun R.E., Carlson C.W., McFadden J.P., Bieging J.H., Delroy G.T., 2000, ApJ, 538, 456
\bibitem[\protect\citeauthoryear{Forbrich \& Berger}{2009}]{forbrich_berger09} 
	Forbrich J., Berger E., 2009, ApJ, 706, L205
	
\bibitem[\protect\citeauthoryear{Forbrich, Berger \& Reid}{Forbrich et al.}{2013}]{forbrich_etal13} 
	Forbrich J., Berger E., Reid M., 2013, ApJ, 777, 70 
	
\bibitem[\protect\citeauthoryear{Gawro\'{n}ski, Go\'{z}dziewski \& Katarzy\'{n}ski}{Gawro\'{n}ski et al.}{2017}]{gawronski_etal16} 
	Gawro\'{n}ski M.P., Go\'{z}dziewski K., Katarzy\'{n}ski K., 2017, MNRAS, 466, 4211 

\bibitem[\protect\citeauthoryear{Gillon et al.}{2017}]{gillon_etal17} 
	Gillon M., Triaud A.H.M.J., Demory B.-O., et al., 2017, Nature, 542, 456	



\bibitem[\protect\citeauthoryear{Gizis et al.}{2016}]{gizis_etal16} 
	Gizis J.E., Williams P.K.G., Burgasser A.J., et al., 2016, AJ, 152, 123	

\bibitem[\protect\citeauthoryear{Glagolevskij}{2011}]{glagolevskij11} 
	Glagolevskij Y.V., 2011, AstBu, 66, 144

\bibitem[\protect\citeauthoryear{Guedel \& Benz}{1993}]{guedel_benz93} 
	Guedel M., Benz A.O., 1993, ApJ, 405, L63
\bibitem[\protect\citeauthoryear{Hallinan et al.}{2006}]{hallinan_etal06} 
	Hallinan G., Antonova A., Doyle J.G., Bourke S., Brisken W.F., Golden A., 2006, ApJ, 653, 690 
\bibitem[\protect\citeauthoryear{Hallinan et al.}{2007}]{hallinan_etal07} 
	Hallinan G., Bourke S., Lane C., et al., 2007, ApJ, 663, L25	
\bibitem[\protect\citeauthoryear{Hallinan et al.}{2008}]{hallinan_etal08} 
	Hallinan G., Antonova A., Doyle J.G., Bourke S., Lane C., Golden A., 2008, ApJ, 684, 644	
\bibitem[\protect\citeauthoryear{Hallinan et al.}{2015}]{hallinan_etal15}
	Hallinan G., Littlefair S.P., Cotter G., et al., 2015, Nature, 523, 568
	% Bourke S., Harding L.K., Pineda J.S., Butler R.P., Golden A., et al., 2015, Nature, 523, 568
	
	
\bibitem[\protect\citeauthoryear{Harding et al.}{2013}]{harding_etal13}
	Harding L.K., Hallinan G., Boyle R.P., Golden A., Singh N., Sheehan B., Zavala R.T., Butler R., 2013, ApJ, 779, 101	

\bibitem[\protect\citeauthoryear{Hatzes}{1997}]{hatzes97}
        Hatzes A.P., 1997, MNRAS, 288, 153

\bibitem[\protect\citeauthoryear{Hess, Cecconi \& Zarka}{Hess et al.}{2008}]{hess_etal08} 
	Hess S., Cecconi B., Zarka P., 2008, GeoRL, 35, L13107

\bibitem[\protect\citeauthoryear{Hess et al.}{2010}]{hess_etal10}
	Hess S.L.G., P\'{e}tin, Zarka P., Bonfond B., Cecconi B., 2010, P\&SS, 58, 1188

\bibitem[\protect\citeauthoryear{Hess \& Zarka}{2011}]{hess_zarka11} 
	Hess S., Zarka P., 2011, A\&A, 531, 29
	
\bibitem[\protect\citeauthoryear{Hess et al.}{2011}]{hess_etal11} 
	Hess S., Bonfond B., Zarka P., Grodent D., 2011, J. Geophys. Res., 116, A05217

\bibitem[\protect\citeauthoryear{Hill et al.}{Hill, Dessler \& Michel}{1974}]{hill_etal74} 
	Hill T.W, Dessler A.J., Michel F.C., 1974, GeoRL, 1, 3

	
\bibitem[\protect\citeauthoryear{Hill}{1979}]{hill_79} 
	Hill T.W, 1979, J. Geophys. Res, 84, 6554

	
\bibitem[\protect\citeauthoryear{Jaeger et al.}{2011}]{jaeger_etal11} 
	Jaeger T.R., Osten R.A., Lazio T.J., Kassim N., Mutel R.L., 2011, AJ, 142, 189	
\bibitem[\protect\citeauthoryear{Kao et al.}{2016}]{kao_etal16} 
	Kao M.M., Hallinan G., Pineda J.S., Escala I., Burgasser A. Bourke S., Stevenson D., 2016, ApJ, 818, 24	
\bibitem[\protect\citeauthoryear{Kuker \& Rudiger}{1999}]{kuker_rudiger99} 
	Kuker M., Rudiger G., 1999, A\&A, 346, 922
\bibitem[\protect\citeauthoryear{Kuznetsov et al.}{2012}]{kuznetsov_etal12} 
	Kuznetsov A.A., Doyle J.G., Yu S., Hallinan G., Antonova A., Golden A., 2012, ApJ, 746, 99
	
	
\bibitem[\protect\citeauthoryear{Lamy et al.}{2008}]{lamy_etal08} 
	Lamy L., Zarka P., Cecconi S., Hess S., Prang\'{e}, 2008, J. Geophys. Res., 113, A10213

\bibitem[\protect\citeauthoryear{Lane et al.}{2007}]{lane_etal07} 
	Lane C., Hallinan G., Zavala R.T., et al., 2007, ApJ, 668, L163
\bibitem[\protect\citeauthoryear{Leone}{1991}]{leone91} 
	Leone F., 1991, A\&A, 252, 198
\bibitem[\protect\citeauthoryear{Leone \& Umana}{1993}]{leone_umana93} 
	Leone F., Umana G., 1993, A\&A, 268, 667
\bibitem[\protect\citeauthoryear{Leone, Trigilio \& Umana}{Leone et al.}{1994}]{leone_etal94} 
	Leone F., Trigilio C., Umana G., 1994, A\&A, 283, 908
\bibitem[\protect\citeauthoryear{Leto et al.}{2006}]{leto_etal06} 
	Leto P., Trigilio C., Buemi C.S., Umana G., Leone F., 2006, A\&A, 458, 831
\bibitem[\protect\citeauthoryear{Leto et al.}{2012}]{leto_etal12} 
	Leto P., Trigilio C., Buemi C.S., Leone F., Umana G.,  2012, MNRAS, 423, 1766
\bibitem[\protect\citeauthoryear{Leto et al.}{2016}]{leto_etal16} 
	Leto P., Trigilio C., Buemi C.S., Umana G., Ingallinera A., Cerrrigone L., 2016, MNRAS, 459, 1159 (Paper I)
	
\bibitem[\protect\citeauthoryear{Leto et al.}{2017}]{leto_etal17} 
	Leto P., Trigilio C., Oskinova L., et al., 2017, MNRAS, 467, 2820 
	%Ignace R., Buemi C.S., Umana G., Ingallinera A., Todt H., Leone F.,
	
\bibitem[\protect\citeauthoryear{Lynch, Mutel \& G\"{u}del}{Lynch et al.}{2015}]{lynch_etal15}
	Lynch C., Mutel R.L., G\"{u}del M., 2015, ApJ, 802, 106
\bibitem[\protect\citeauthoryear{Lynch et al.}{2016}]{lynch_etal16}
	Lynch C., Murphy T., Ravi V., Hobbs G., Lo K., Ward C., 2016, MNRAS, 457, 1224
\bibitem[\protect\citeauthoryear{Linsky, Drake \& Bastian}{Linsky et al.}{1992}]{linsky_etal92} 
	Linsky J.L., Drake S.A., Bastian S.A., 1992, ApJ, 393, 341
\bibitem[\protect\citeauthoryear{Littlefair et al.}{2008}]{littlefair_etal08} 
	Littlefair S.P., Dhillon V.S., Marsh T.R., Shahbaz T., Mart'n E.L., Copperwheat C., 2008, MNRAS, 391, L88
\bibitem[\protect\citeauthoryear{Lo et al.}{2012}]{lo_etal12} 
	Lo K.K., Bray J.D., Hobbs G., et al., 2012, MNRAS, 421, 3316
\bibitem[\protect\citeauthoryear{Louarn \& Le Queau}{1996a}]{louarn_lequeau96a} 
	Louarn P., Le Queau D., 1996a, P\&SS, 44, 199
\bibitem[\protect\citeauthoryear{Louarn \& Le Queau}{1996b}]{louarn_lequeau96b} 
	Louarn P., Le Queau D., 1996b, P\&SS, 44, 211
\bibitem[\protect\citeauthoryear{McLean et al.}{2011}]{mclean_etal11} 
	McLean M., Berger E., Irwin J., Forbrich J., Reiners A., 2011, ApJ, 741, 27	
\bibitem[\protect\citeauthoryear{McLean, Berger \& Reiners}{McLean et al.}{2012}]{mclean_etal12} 
	McLean M., Berger E., Reiners A., 2012, ApJ, 746, 23
\bibitem[\protect\citeauthoryear{Melrose \& Dulk}{1982}]{melrose_dulk82} 
	Melrose D.B., Dulk G.A., 1982, ApJ, 259, 844
\bibitem[\protect\citeauthoryear{Menietti et al.}{2011}]{menietti_etal11} 
	Menietti J.D., Mutel R.L., Christopher I.W., Hutchinson K.A., Sigwarth J.B, 2011, J. Geophys. Res., 116, A12219 
	
\bibitem[\protect\citeauthoryear{Metodieva et al.}{2017}]{metodieva_etal17} 
	Metodieva Y.T., Kuznetsov A.A., Antonova A.E., Doyle J.G., Ramsay G., Wu K., 2017, MNRAS, 465, 1995
	
\bibitem[\protect\citeauthoryear{Mohanty \& Basri}{2003}]{mohanty_basri03} 
	Mohanty S., Basri G., 2003, ApJ, 583, 451
\bibitem[\protect\citeauthoryear{Morin et al.}{2008a}]{morin_etal08a} 
	Morin J., Donati J.-F., Forville T., et al., 2008a, MNRAS, 384, 77	
\bibitem[\protect\citeauthoryear{Morin et al.}{2008b}]{morin_etal08b} 
	Morin J., Donati J.-F., Petit P., et al., 2008b, MNRAS, 390, 567
\bibitem[\protect\citeauthoryear{Morin et al.}{2010}]{morin_etal10} 
	Morin J., Donati J.-F., Petit P., et al., 2010, MNRAS, 407, 2269
\bibitem[\protect\citeauthoryear{Mutel, Christopher \& Pickett}{Mutel et al.}{2008}]{mutel_etal08}
	Mutel R.L., Christopher I.W., Pickett J.S., 2008, GeoRL, 35, L07104.
\bibitem[\protect\citeauthoryear{Nichols}{2011}]{nichols11}
 	Nichols J.D., 2011, MNRAS, 414, 2125
\bibitem[\protect\citeauthoryear{Nichols et al.}{2012}]{nichols_etal12}
 	Nichols J.D., Burleigh M.R., Casewell S.L., Cowley S.W.H., Wynn G.A., Clarke J.T.,  Westand A.A., 2012, ApJ, 760, 59
\bibitem[\protect\citeauthoryear{Nichols \& Milan}{2016}]{nichols_milan16}
 	Nichols J.D., Milan S.E., 2016, MNRAS, 461, 2353
	
\bibitem[\protect\citeauthoryear{Oksala et al.}{2012}]{oksala_etal12} 
	Oksala M.E., Wade G.A., Townsend R.H.D., et al., 2012, MNRAS, 419, 959
\bibitem[\protect\citeauthoryear{Oksala et al.}{2015}]{oksala_etal15} 
	Oksala M.E., Kochukhov O., Krticka J., et al., 2015, MNRAS, 451, 2015

\bibitem[\protect\citeauthoryear{Osten et al.}{2006}]{osten_etal06} 
	Osten R.A., Hawley S.L., Bastian T.S., Reid I.N., 2006, ApJ, 637, 518	

\bibitem[\protect\citeauthoryear{Queinnec \& Zarke}{1998}]{queinnec_zarka98}
	Queinnec J., Zarka P., 1998, J. Geophys. Res., 103, A11	
	
	
\bibitem[\protect\citeauthoryear{Ray \& Hess}{2008}]{ray_hess08} 
	Ray L.C., Hess S., 2008, J. Geophys. Res., 113, A11218
	
	
\bibitem[\protect\citeauthoryear{Ravi et al.}{2010}]{ravi_etal10} 
	Ravi V., Hobbs G., Wickramasinghe D., Champion D.J., Keith M., 2010, MNRAS, 408, L99
\bibitem[\protect\citeauthoryear{Reiners \& Basri}{2007}]{reiners_basri07} 
	Reiners A., Basri G., 2007, ApJ, 656, 1121
\bibitem[\protect\citeauthoryear{Reiners \& Basri}{2010}]{reiners_basri10} 
	Reiners A., Basri G., 2010, ApJ, 710, 924
\bibitem[\protect\citeauthoryear{Ricci et al.}{2012}]{ricci_etal12} 
	Ricci L., Testi L., Natta A., Scholz A., de Gregorio-Monsalvo I., 2012, ApJ, 761, L20
\bibitem[\protect\citeauthoryear{Ricci et al.}{2013}]{ricci_etal13} 
	Ricci L., Isella A., Carpenter J.M., Testi L., 2013, ApJ, 764, L27	
\bibitem[\protect\citeauthoryear{Route \& Wolszczan}{2012}]{route_wolszczan12} 
	Route M., Wolszczan A., 2012, ApJ, 747, L22	
\bibitem[\protect\citeauthoryear{Route \& Wolszczan}{2013}]{route_wolszczan13}
	Route M., Wolszczan A., 2013, ApJ, 773, 18
\bibitem[\protect\citeauthoryear{Route \& Wolszczan}{2016}]{route_wolszczan16}
	Route M., Wolszczan A., 2016, ApJ, 821, L21
	
	
\bibitem[\protect\citeauthoryear{Route}{2016}]{route16}
	Route M., 2016, ApJ, 830, L27
\bibitem[\protect\citeauthoryear{Schmidt et al.}{2007}]{schmidt_etal07} 
	Schmidt S.J.,  Cruz K.L., Bongiorno B.J., Liebert J., Reid I.N., 2007, ApJ, 133, 2258
\bibitem[\protect\citeauthoryear{Schrijver}{2009}]{schrijver_09} 
	Schrijver C.J., 2009, ApJ, 773, L148
\bibitem[\protect\citeauthoryear{Speir et al.}{2014}]{speir_etal14} 
	Speir D.C., Bingham R., Cairns R.A., Vorgul I., Kellett B.J., Phelps A.D.R., Ronald K., 2014, Physical Review Letters, 113, 155002

\bibitem[\protect\citeauthoryear{Tingay et al.}{2013}]{tingay_etal13} 	
	Tingay S.J., Goeke R., Bowman J.D., et al., 2013, PASA, 30, e007
\bibitem[\protect\citeauthoryear{Trigilio et al.}{2000}]{trigilio_etal00} 
	Trigilio C., Leto P., Leone F., Umana G., Buemi C., 2000, A\&A, 362, 281
\bibitem[\protect\citeauthoryear{Trigilio et al.}{2004}]{trigilio_etal04} 
	Trigilio C., Leto P., Umana G., Leone F., Buemi C.S., 2004, A\&A, 418, 593
\bibitem[\protect\citeauthoryear{Trigilio et al.}{2008}]{trigilio_etal08} 
	Trigilio C., Leto P., Umana G., Buemi C.S., Leone F., 2008, MNRAS, 384, 1437
\bibitem[\protect\citeauthoryear{Trigilio et al.}{2011}]{trigilio_etal11} 
	Trigilio C., Leto P., Umana G., Buemi C.S., Leone F., 2011,  ApJ, 739, L10
	
\bibitem[\protect\citeauthoryear{van Haarlem et al.}{2013}]{van_haarlem_etal13} 
	van Haarlem M.P., Wise M.W., Gunst A.W., et al., 2013, A\&A, 556, A2
	
\bibitem[\protect\citeauthoryear{Vorgul \& Helling}{2016}]{vorgul_helling16}
	Vorgul I., Helling Ch., 2016, MNRAS, 458, 104
\bibitem[\protect\citeauthoryear{Williams, Cook \& Berger}{Williams et al.}{2014}]{williams_etal14}
	Williams P.K.G., Cook B.A., Berger E.,  2014, ApJ, 785, 9
\bibitem[\protect\citeauthoryear{Williams et al.}{2015a}]{williams_etal15a}
	Williams P.K.G., Berger E., Irwin J., Berta-Thompson Z.K., Charbonneau D.,  2015, ApJ, 799, 192
\bibitem[\protect\citeauthoryear{Williams et al.}{2015b}]{williams_etal15b}
	Williams P.K.G., Casewell S.L.,  Stark C.R., Littlefair  S.P., Helling Ch., Berger E.,  2015, ApJ, 815, 64
\bibitem[\protect\citeauthoryear{Williams \& Berger}{2015}]{william_berger15}
	Williams P.K.G., Berger E., 2015, ApJ, 808,189
	
\bibitem[\protect\citeauthoryear{Williams, Gizis \& Berger}{Williams et al.}{2017}]{williams_etal17}
	Williams P.K.G., Gizis J.E., Berger E.,  2017, ApJ, 834, 117

	
\bibitem[\protect\citeauthoryear{Winglee \& Pritchett}{1986}]{winglee_pritchett86} 
	Winglee R.M., Pritchett P.L., 1986, J. Geophys. Res., 91, 13531
\bibitem[\protect\citeauthoryear{Wolszczan \& Route}{2014}]{wolszcan_route14}
	Wolszczan A., Route M., 2014, ApJ, 788, 23
\bibitem[\protect\citeauthoryear{Wu \& Lee}{1979}]{wu_lee79} 
	Wu C.S., Lee L.C., 1979, ApJ, 230, 621
	
\bibitem[\protect\citeauthoryear{Yadav et al.}{2015}]{yadav_etal15} 
	Yadav R.K., Christensen U.R., Morin J., Gastine T., Reiners A., Poppenhaeger K., Scott J.W., 2015, ApJ, 813, L31

\bibitem[\protect\citeauthoryear{Yadav et al.}{2016}]{yadav_etal16} 
	Yadav R.K., Christensen U.R., Scott J.W., Poppenhaeger K., 2016, ApJ, 833, L28
	
	
\bibitem[\protect\citeauthoryear{Yoneda et al.}{2013}]{yoneda_etal13} 
	Yoneda M., Tsuchiya  F.,  Misawa H., Bonfond C., Tao C., Kagitani M., Okano S., 2013, GeoRL, 40, 671
\bibitem[\protect\citeauthoryear{Yu et al.}{2011}]{yu_etal11} 
	Yu S., Hallinan G., Doyle J.G., et al., 2011, A\&A, 525, A39
\bibitem[\protect\citeauthoryear{Zarka}{1998}]{zarka98} 
	Zarka P., 1998, J. Geophys. Res., 103, 20159
\bibitem[\protect\citeauthoryear{Zarka}{2007}]{zarka07} 
	Zarka P., 2007, P\&SS, 55, 598
\end{thebibliography}
\end{document}